\RequirePackage{fix-cm}
\documentclass[onecolumn]{svjour3}
\smartqed 
\usepackage[utf8]{inputenc}          %
\usepackage{graphicx}
\usepackage{xcolor}
\usepackage{listings}
\definecolor{dkgreen}{rgb}{0,0.6,0}
\usepackage{booktabs}
\usepackage{siunitx}
\usepackage{subcaption}
\usepackage{graphicx}
\usepackage{amsmath}
\usepackage[export]{adjustbox}
\usepackage{pythonhighlight}
\usepackage{caption}
\usepackage{multirow}

\usepackage[round]{natbib}
\usepackage[hyphens]{url}	

\usepackage{xurl}

\usepackage{float}

\usepackage{setspace}
\makeatletter
\let\cl@chapter\undefined
\makeatletter
\usepackage[capitalise]{cleveref}
\usepackage[margin=1in,includefoot,footskip=30pt]{geometry}

\usepackage{lineno}

\begin{document}

\title{An inverse analysis of fluid flow through granular media using differentiable lattice Boltzmann method}

\author{Qiuyu Wang\textsuperscript{*} \and Krishna Kumar}

\institute{
      Q. Wang, K. Kumar \at
The University of Texas at Austin \\
Civil, Architectural and Environmental Engineering \\
Cockrell School of Engineering \\
301 E. Dean Keeton St. C2100, Austin, Texas 78712-2100, USA\\
\email{wangqiuyu@utexas.edu}
}

\newcommand{\ie}{\textit{i.\,e.},~}
\newcommand{\eg}{\textit{e.\,g.},~}
\newcommand{\etc}{\textit{etc.}}
\date{}
\maketitle
\doublespacing

\begin{abstract}
In this study, we introduce an effective method for the inverse analysis of fluid flow problems, focusing on accurately determining boundary conditions and characterizing the physical properties of granular media, such as permeability, and fluid components, like viscosity. Our primary aim is to deduce either constant pressure head or pressure profiles, given the known velocity field at a steady-state flow through a conduit containing obstacles, including walls, spheres, and grains. We employ the lattice Boltzmann Method (LBM) combined with Automatic Differentiation (AD), facilitated by the GPU-capable Taichi programming language (AD-LBM). A lightweight tape is utilized to generate gradients for the entire LBM simulation, enabling end-to-end backpropagation. For complex flow paths in porous media, our AD-LBM approach accurately estimates the boundary conditions leading to observed steady-state velocity fields and consequently derives macro-scale permeability and fluid viscosity. Our method demonstrates significant advantages in terms of prediction accuracy and computational efficiency, offering a powerful tool for solving inverse fluid flow problems in various applications.

\end{abstract}

\section{Introduction}
\label{sec:intro}

Accurate modeling of fluid flow through porous granular media is critical for diverse engineering applications, including hydrocarbon recovery optimization, grouts and filter design, and stability analysis against seepage. However, inverse prediction of the ideal boundary conditions or fluid properties to achieve a desired flow state remains challenging. Furthermore, this inverse problem is ill-posed and sensitive to measurement noise~\citep{cheylan2019shape}. Solving these inverse problems involves calculating functional gradients of the specified flow conditions (results) on a set of fluid state variables (inputs) and then minimizing the gradients to find optimal solutions~\citep{plessix2006review}. However, traditional CFD tools are limited to forward differentiation and cannot efficiently calculate these reverse-mode gradients~\citep{norgaard2017applications}. Specialized techniques like adjoint solvers, which require manually defining the gradient function of the forward problem or finite difference (FD) approximations, require extensive computations and memory for gradient estimation in time-dependent problems~\citep{cheylan2019shape,tekitek2006adjoint,rokicki2016adjoint}. FD gradient computation requires multiple runs and may suffer from numerical instabilities or experience vanishing gradients for small pertubations. Hence, there is a need for more specialized techniques to compute multi-parameter gradients for solving the optimization of fluid flow through granular media.

Differentiable simulations combine automatic differentiation with gradient-based optimization to build a simulation framework that can compute derivatives relating output to its input~\citep{battaglia2018relational,de2019deep}. Treating simulation as a differentiable function facilitates parameter optimization and control. Moreover, unlike numerical approximations, automatic differentiation ensures accurate and efficient gradient computations. This combination of automatic differentiation with fluid simulation paves the way for integrating physical simulations with machine learning algorithms, opening up exciting avenues for research and applications in physics-informed machine learning and data-driven science.

Differentiable simulation has significant implications for fluid flow through granular media, where it can provide a powerful tool for inverse problems and control tasks. For example, it can estimate the forces or boundary conditions that resulted in observed fluid flow or determine control inputs that would cause a fluid to move in a desired way. Despite the advancements, auto differentiation has limited applications in optimizing flow dynamics for shape and topology~\citep{mohammadi2009applied,rozvany2014topology}. JAX-FLUIDS is another notable example of a fully differentiable high-order computational fluid dynamics code that leverages machine learning and differential programming to solve Navier-Stokes equations and simulate turbulence~\citep{bezgin2023jax}.

This work presents a novel differentiable lattice Boltzmann model (LBM) for efficient inversion of fluid flows through granular media. Leveraging the automatic differentiation capabilities of the Taichi programming language, we develop an innovative framework to backpropagate through LBM simulations. We successfully demonstrate the determination of pressure boundary conditions from limited velocity data. Despite having access to only partial velocity fields, our approach can deduce full pressure profiles to within 10\% average error. Furthermore, we showcase the ability to evaluate permeability and identify fluid viscosity matching observed flows within 4\% error. Our differentiable LBM represents the first demonstrated method for optimizing complex granular media flows. This paradigm paves the way for integrating physical simulations with machine learning toward data-driven modeling of subsurface flow. While we focus on laminar steady-state cases, future efforts will aim to extend this approach to turbulent and transient flows.

\section{Methodology}
\subsection{Lattice Boltzmann Method}
\label{sec:physical_model}

The fluid state at any point x =[x, y, z]$^T$  and time t is governed by the Navier-Stokes equations, which describe momentum and mass conservation for fluid flow. However, solving continuous partial differential equations numerically can be computationally demanding. The LBM offers an alternative mesoscopic approach based on kinetic theory models of fluids~\citep{kruger2017lattice}.

In LBM, fluid dynamics emerge from collective behaviors of virtual particle distributions on a lattice. LBM solves a discretized version of the Boltzmann equation by locally streaming and colliding discrete particle populations. LBM can simulate macroscopic fluid flows efficiently while inherently including mesoscale interactions. The local nature of LBM is well-suited for coupling fluid dynamics with granular dynamics.

\begin{figure}[htbp]
    \centering
    \begin{subfigure}[b]{0.31\textwidth}
        \includegraphics[width=\textwidth]{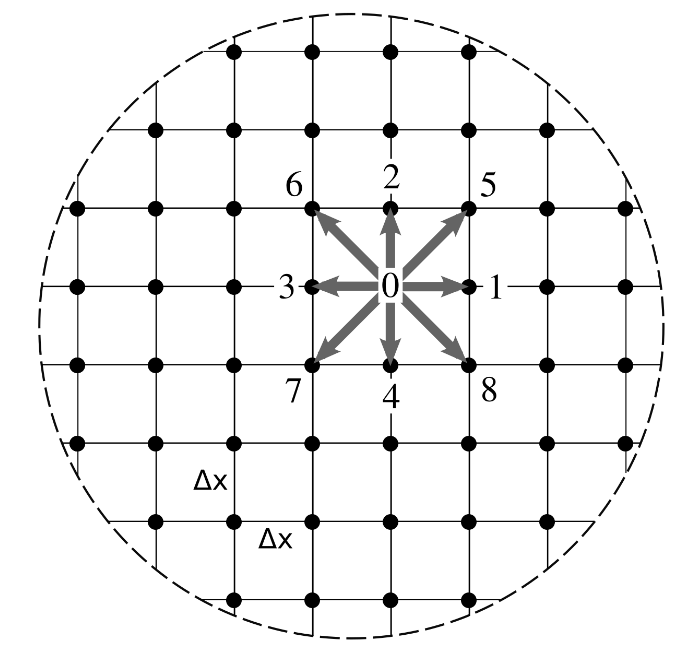}
        \caption{D2Q9}
        \label{fig:d2q9}
    \end{subfigure}
    \hfill
    \begin{subfigure}[b]{0.31\textwidth}
        \includegraphics[width=\textwidth]{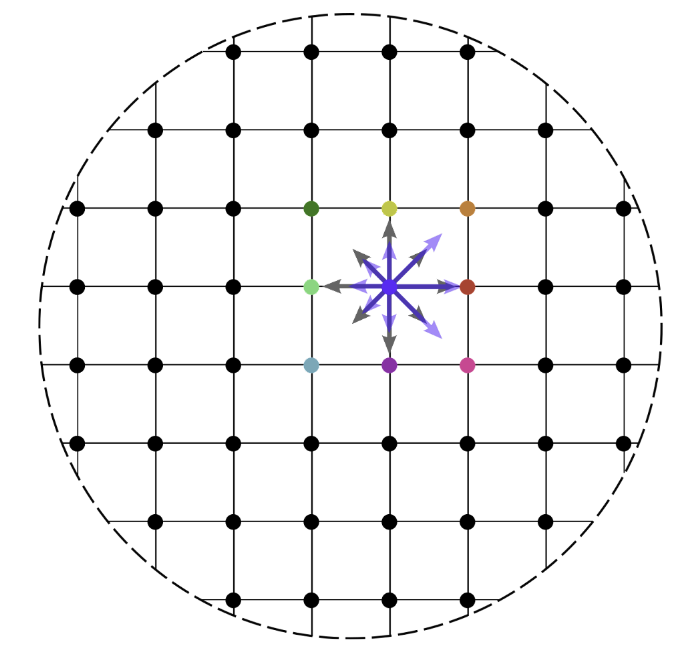}
        \caption{Collision}
        \label{fig:collision}
    \end{subfigure}
    \hfill
    \begin{subfigure}[b]{0.31\textwidth}
        \includegraphics[width=\textwidth]{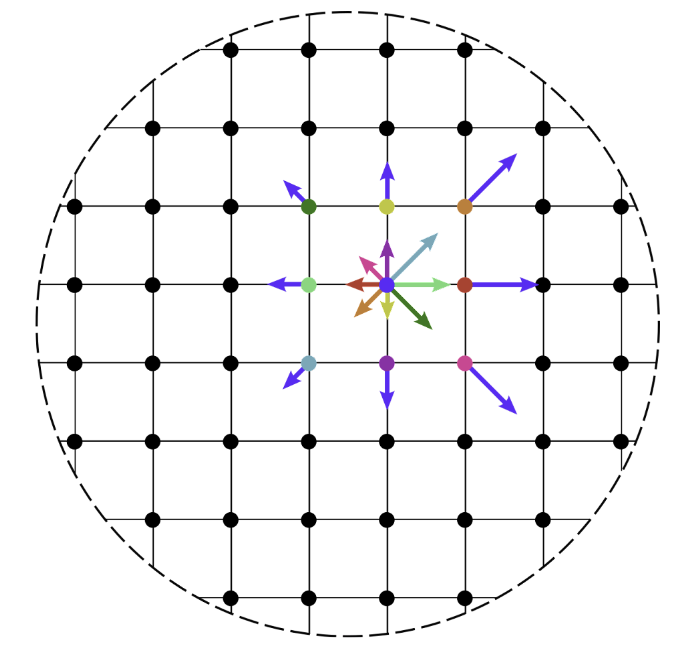}
        \caption{Streaming}
        \label{fig:streaming}
    \end{subfigure}
    \caption{D2Q9 velocity sets. The square denoted by solid lines has an edge length dx.}
    \label{fig:D2Q9}
\end{figure}

We use the D2Q9 scheme for 2D modeling and D3Q19 for 3D modeling. The number of discrete velocities is denoted by Q.~\Cref{fig:D2Q9} shows the D2Q9 velocity sets as well as the collision and streaming stages in LBM simulation. As seen in~\Cref{fig:d2q9}, the lattice was created by dividing the fluid domain into an equally spaced grid in 9 or 19 directions, and each intersection of the grid was referred to as a lattice node. The particle distribution function, $f_i$, was associated with each lattice node at each discrete direction, i, which represented the density of particles with velocity $c_i$ at that node. 

\begin{equation}
\rho=\sum_{i=0}^{19}f_{i},\,
\end{equation}
Macroscopic quantities such as mass density, rho, or fluid velocity, $u$, which are usually of interest in fluid dynamics, could be calculated from the distribution functions using the lattice Boltzmann equation:
\begin{equation}
\rho\boldsymbol{u}=\sum_{i=0}^{19}f_{i}\boldsymbol{c}_{i},\,
\end{equation}
Following the definition of the velocities, an evolution rule is applied to solve the Boltzmann equation at every time increment:
\begin{equation}
f_{i}(\boldsymbol{x}+\boldsymbol{c}_{i}\Delta t,t+\Delta t)=f_{i}(\boldsymbol{x},t)+\mathrm{\Omega}_{col},\,
\end{equation}
where $\boldsymbol{x}$ is the position of the lattice node, $t$ is the current time and $\mathrm{\Omega}_{col}$ is the Bhatnagar–Gross–Krook (BGK) collision operator~\citep{d1986lattice}, corresponding to~\Cref{fig:collision}:
\begin{equation}
\mathrm{\Omega}_{col}=\frac{f_{i}-f_{i}^{eq}}{\tau }\Delta t,\,
\label{eq:bgk_operator1}
\end{equation}
During collision, the collision operator depends on the relaxation time $\tau$ which we use ${\tau = \Delta t}$ in our simulations. The equilibrium distribution function, $f_{i}^{eq}$ , in~\cref{eq:bgk_operator1} is given by,
\begin{equation}
f_{i}^{eq}=\omega _{i}\rho (1+\frac{\boldsymbol{u}\cdot \boldsymbol{c}_{i}}{\boldsymbol{c}_{s}^{2}}+\frac{(\boldsymbol{u}\cdot \boldsymbol{c}_{i})^2}{2\boldsymbol{c}_{s}^{4}}-\frac{\boldsymbol{u}\cdot \boldsymbol{u}}{2\boldsymbol{c}_{s}^{2}}),\,
\end{equation}
where $\omega _i$ as the corresponding weights. After a collision, the updated distribution functions are streamed according on their velocities: each $f_i$ is pushed one lattice node in the $i$ direction, as shown in~\Cref{fig:streaming}.

To simulate the fluid flow driven by a constant pressure difference between the inlet and outlet faces, we employ the anti-bounced back conditions at boundary~\citep{ginzburg2008two}:
\begin{equation}
f_{i}(x_b,t+\Delta t)=-f_{i}(x_b,t)+2\omega _{i}\rho_w (1+\frac{(\boldsymbol{u}\cdot \boldsymbol{c}_{i})^2}{2\boldsymbol{c}_{s}^{4}}-\frac{\boldsymbol{u}\cdot \boldsymbol{u}}{2\boldsymbol{c}_{s}^{2}}),\,
\end{equation}
where $\rho_w$ is the constant fluid density of the boundary nodes ($x=x_b$) leading to a constant fluid pressure. 

In LBM simulations, the parameters such as grid spacing (dx), timestep (dt), and relaxation time ($\tau$) are set to 1 in dimensionless lattice units. Establishing geometric and dynamic similarity between physical and lattice scales is a crucial step, which is achieved by matching dimensionless parameters like Reynolds number ($R_e$), Rayleigh number, Nusselt number, and Weber number, among others. Regardless of fluid velocity, geometric scale, or viscosity differences, flows with identical Reynolds numbers are deemed comparable. These dimensionless numbers allow the derivation of physical properties from LBM simulations, such as permeability, thermal conductivity, and formation factor. The accuracy of the numerical solutions for these physical properties from LBM simulations can be relied upon, even in cases of significant geometric complexity. This process applies to LBM simulations employing a BGK collision operator ~\citep{huang2015multiphase,baakeem2021novel}.

\subsection{Differentiable Lattice Boltzmann Simulation}
\label{sec:Numerical_model}

\begin{figure}[htbp]
  \begin{subfigure}{0.9\columnwidth}
    \includegraphics[width=0.9\columnwidth]{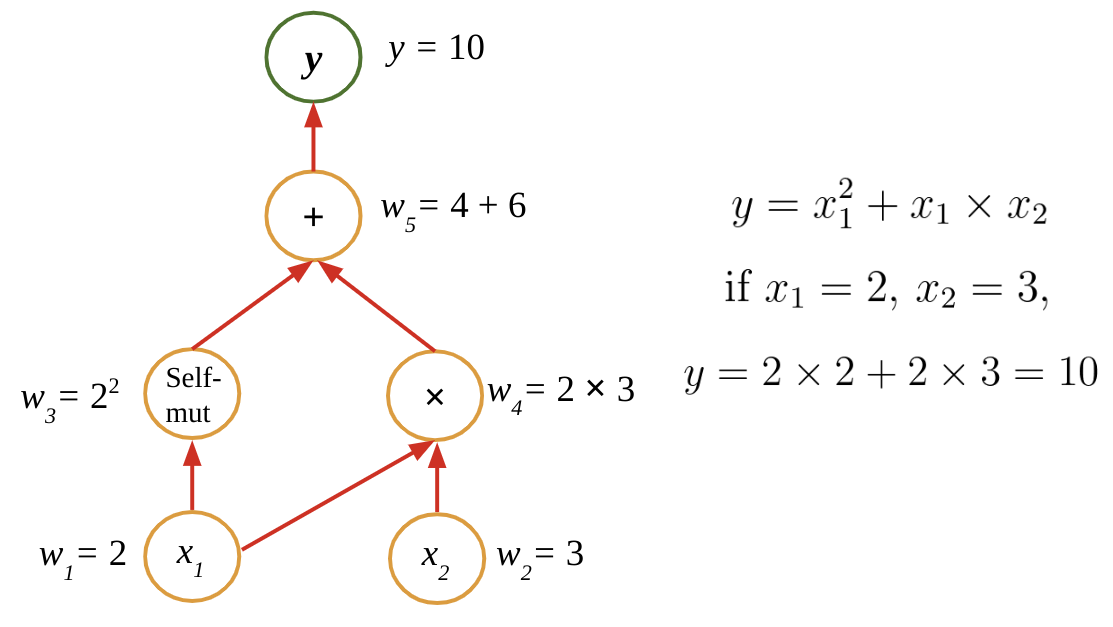}
    \caption{Forward mode}
  \end{subfigure}
  \begin{subfigure}{0.9\columnwidth}
    \includegraphics[width=0.9\columnwidth]{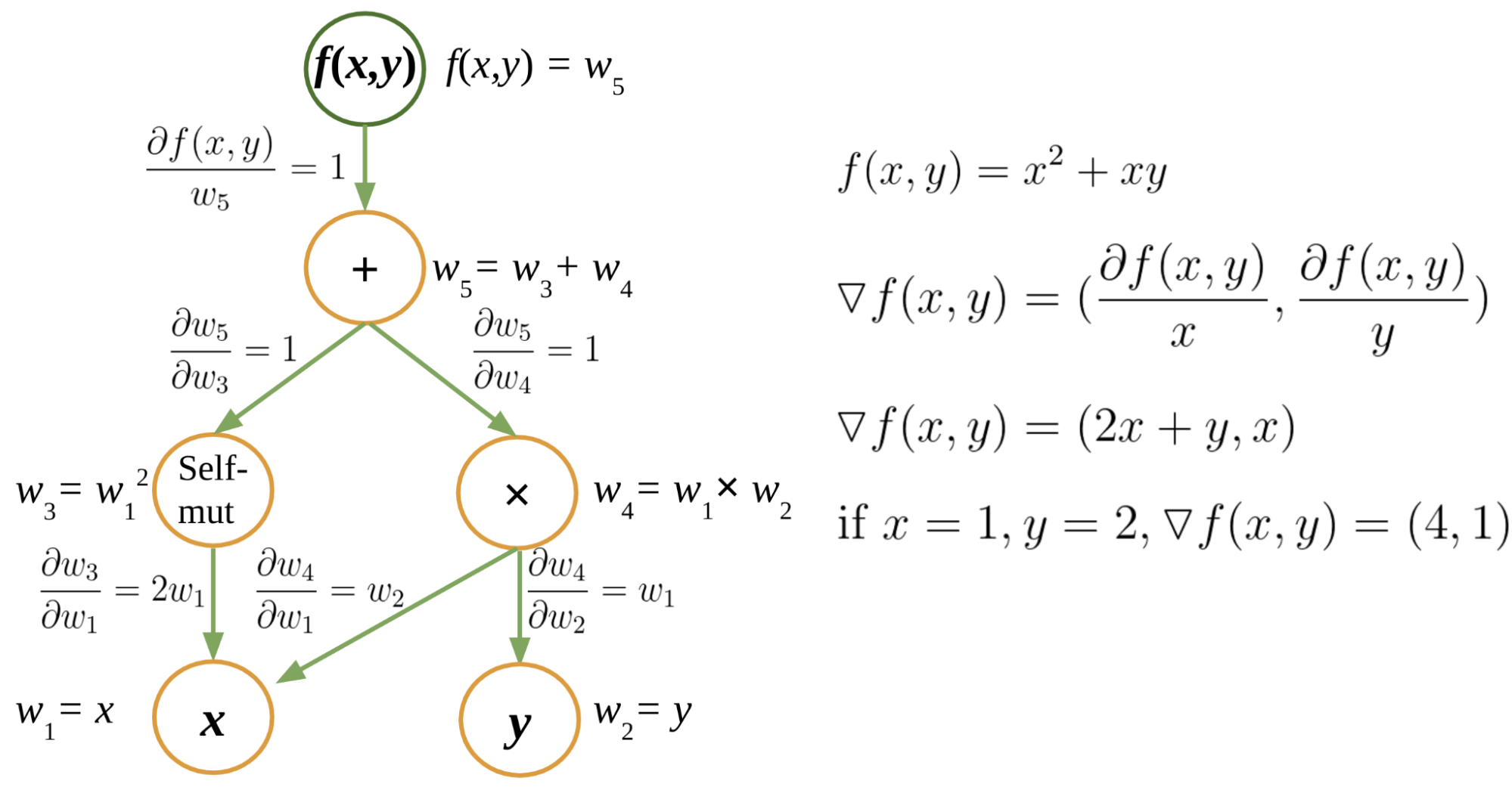}
    \caption{Reverse mode}
  \end{subfigure}%
\caption{Directed Acyclic Graph (DAG) of $f(x,y)=x^2+xy$.}
\label{Fig:DAG}
\end{figure}

For computing the gradients in simulations, there are four possible methods~\citep{gunes2015automatic,margossian2019review}: (1) manual differentiation; (2) finite difference or other numerical approximations; (3) symbolic differentiation; and (4)  automatic differentiation  (AD). Manual differentiation is time-consuming and impractical for complex with many parameters. Numerical differentiation involves computing the finite difference approximation of derivatives by incorporating a tiny increment to the original function. The finite difference approximation is inaccurate because of truncation and round-off errors~\citep{jerrell1997automatic,burden2015numerical}. Symbolic differentiation calculates the derivative of a mathematical function using its symbolic representation, providing a precise solution without approximations. However, it can result in the problem of "expression swell," yielding complex and lengthy expressions~\citep{corliss1988applications}. Automatic differentiation (AD) combines the power of symbolic and numerical differentiation by breaking down numerical computations into simple differentiable operators, allowing for efficient and accurate computation of derivatives. Hence, AD is a powerful candidate for solving gradients to infer unknown parameters in fluid flow problems. 

We develop a fully differentiable 2D/3D LBM solver using the domain-specific Taichi programming language~\citep{hu2019difftaichi}. To tackle the differential operators inherent in the inverse simulation of Lattice Boltzmann equations (reverse-mode differentiation), we employ automatic differentiation, which delivers substantial improvements in computational efficiency compared to numerical differentiation methods~\citep{gunes2015automatic}. AD cleverly circumvents the drawbacks of numerical and symbolic differentiation by constructing a computational Directed Acyclic Graph (DAG), where nodes represent operations or variables, and edges signify the dependency between them. The flow of this graph captures the sequence of elemental functions and their derivative calculations, providing a precise mechanism for gradient calculation. 

\Cref{Fig:DAG} shows an example of forward and reverse mode automatic differentiation using a Directed Acyclic Graph (DAG). Consider the function \( f(x,y) = x^2 + x \times y \). In the forward mode of AD, computations proceed from the input variables to the output function, following the topological ordering of the graph, as shown in~\cref{Fig:DAG}a. For example, if \( x = 1 \) and \( y = 2 \), then \( f(x,y) = 1^2 + 1 \times 2 \). In the reverse mode AD, often known as backpropagation in neural networks. Computations proceed in the opposite direction, from the output function back to the inputs, as illustrated in~\cref{Fig:DAG}b. At each node, the gradient, represented by \( \nabla \), is calculated with respect to the input variable. To compute the gradients with respect to the input, we perform a chain rule traversing every node from the output to the input node. Using the same example with \( x = 1 \) and \( y = 2 \), the gradients are \( \nabla f(x,y) = (4,1) \). This mode is especially useful when the function has many inputs and few outputs. It computes the derivatives of a single output with respect to all inputs in a single pass. As the computational DAG is traversed, nodes are updated using the chain rule of differentiation, ensuring accurate and efficient gradient calculations. Hence, forward and reverse mode ADs provide a flexible, robust, and efficient framework for calculating derivatives in numerical simulations.

\begin{figure}
    \centering
    \includegraphics[width=\textwidth]{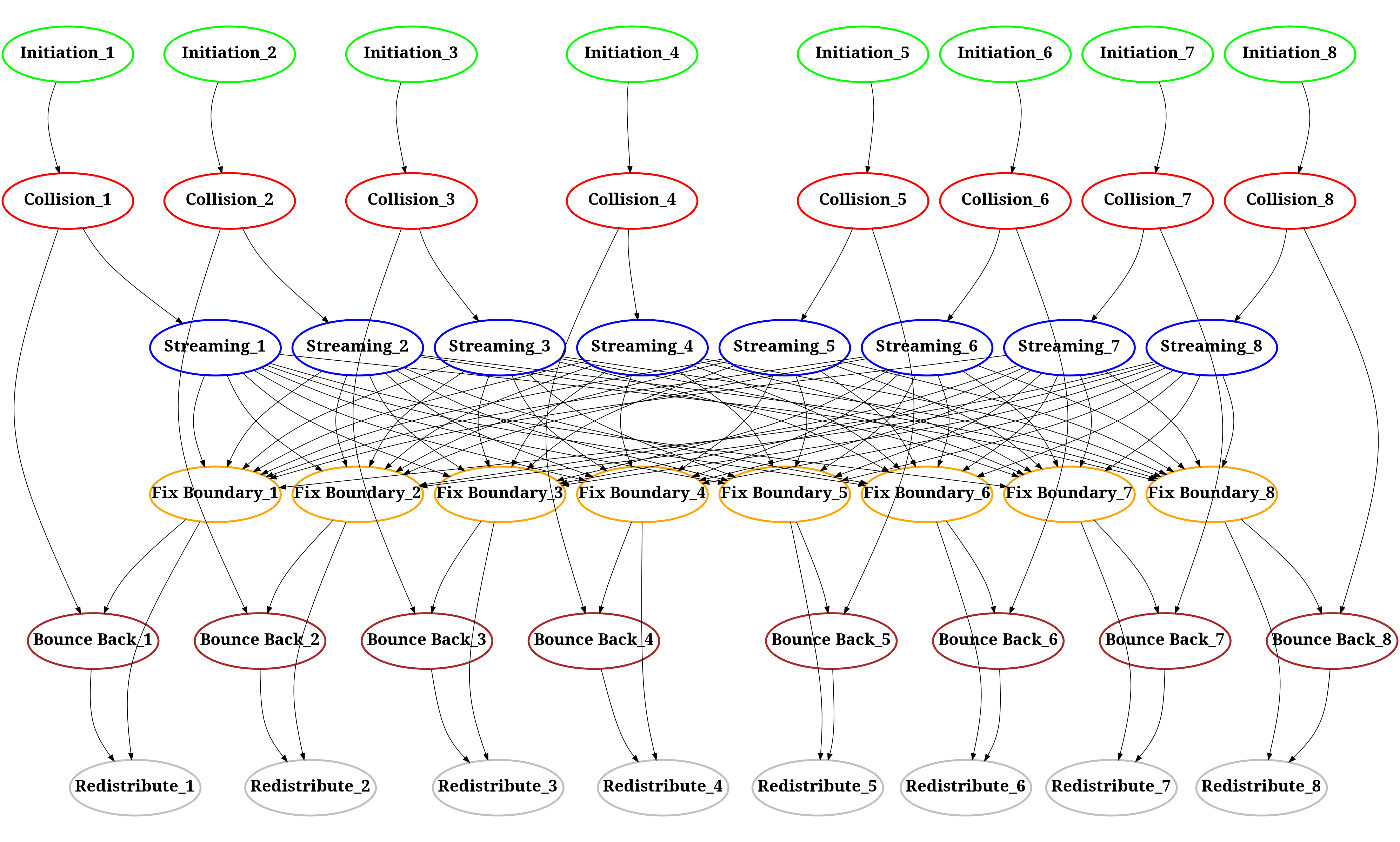}
    \caption{Directed Acyclic Graph (DAG) Representation of lattice Boltzmann Method Simulation Stages with two time steps and 2×2×2 domain division.}
    \label{Fig:3Pipelines}
\end{figure}

Similar to a DAG-based AD in~\cref{Fig:DAG}, we construct a computational DAG for the LBM simulation.~\Cref{Fig:3Pipelines} shows the computational DAG representing a single D2Q9 LBM cell. Each stage of the LBM process—"Initialization," "Collision," "Streaming," "Boundary Fixation," "Bounce Back," and "Redistribution"—is represented as a node in the DAG, and the directed connections between these nodes represent the computational flow.

We represent the AD-LBM process as a high-level abstraction known as the Intermediate Representation (IR) graph. The IR graph operates within compilers as a transitional platform between the source and target machine codes, which the computer's hardware executes. The Taichi programming language constructs this IR graph by monitoring the invoked kernels and displaying the computations of the forward kernels with respect to the inputs~\citep{hu2019difftaichi}. We traverse the forward IR graph in reverse order to generate reverse mode gradients and incorporate appropriate gradient computations for each active function (node) in the forward mode computational graph~\citep{griewank2008evaluating}. A lightweight tape gradient is employed to log the forward kernel launches and replay the gradient kernels in reverse order for backpropagation ~\citep{hu2019taichi}.

\begin{figure}[htbp]
    \centering
    \includegraphics[width=0.95\columnwidth]{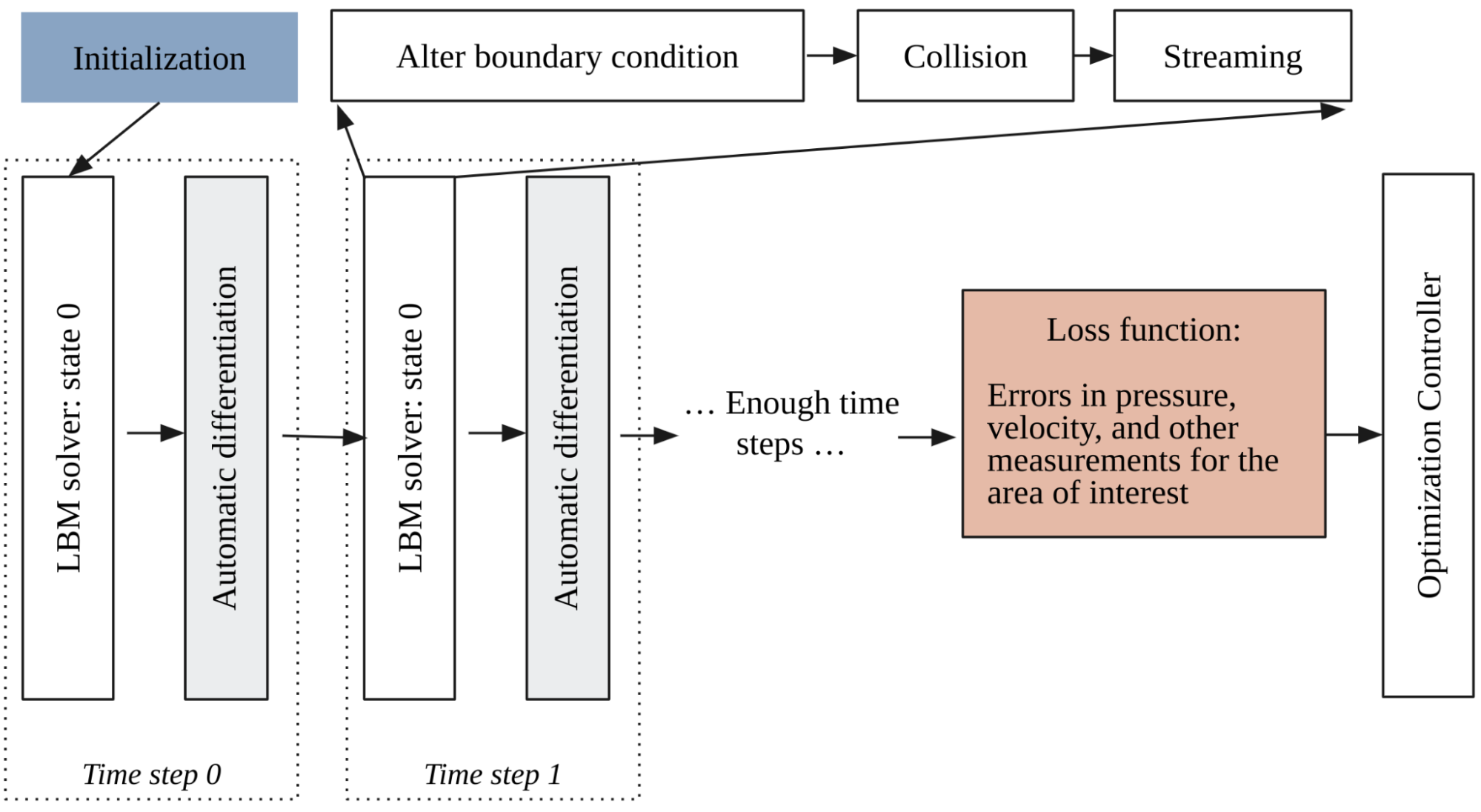}
    \caption{Flowchart of LBM simulation of fluid flow employing Taichi differentiable simulators.}
    \label{Fig:4Flowchart}
\end{figure}

Finally, we combine the AD-LBM solver with a gradient-based optimization approach to build a differentiable simulator.~\Cref{Fig:4Flowchart} shows the workflow of a differentiable LBM simulation capable of solving inverse or design problems involving fluid flow. The LBM differentiable simulator consists of two stages: computing the gradients using automatic differentiation in the reverse mode and a gradient-based optimizer. The differentiable LBM simulation has three stages: initialization, time iterations, and loss function computation. In each time step of the LBM solver, we use AD to compute the time and spatial derivatives for the objective field(s) based on the loss function and accumulate the gradients through the simulation. 

We use gradient-based optimizers, Stochastic Gradient Descent, and ADAM, to solve inverse problems. Consider the problem of identifying the pressure boundary conditions that cause a particular velocity field distribution in a granular media. In this inverse problem, we iteratively optimize the boundary pressure by minimizing the loss function on the known velocity distribution. The loss function is computed as the sum of the squared differences between the simulated velocity norm at each fluid node and the observed values at the target time step. The core kernel function for differentiable simulation to evaluate the loss based on the velocity prediction is shown below:

\begin{lstlisting}
@ti.kernel
def compute_loss():  
    # Loop over the velocity field
    for I in ti.grouped(v): 
        loss[None]+= (v[I].norm()-target_v[I].norm())**2
# "loss" is a scalar (0-D tensor) and indexed by [None].
\end{lstlisting}

To calculate the boundary condition of the input pressure profile, we use the tape gradient in the LBM differential simulator `ti.ad.Tape()` on the kernel function to compute the gradient of the inlet pressure profile as inlet\_pressure.grad[None]($=\frac{\partial loss}{\partial inlet\_pressure}$). `

\begin{lstlisting}
learning_rate =  0.01
# Initial guess of unknown pressure at inlet
loss[None] = 0.
inlet_pressure[None] = 0.6
for  iteration in range(100):
    # Initialize simulation
    init_field()
    # Run simulation for 5000 time steps
    for step in range(5000):
        with ti.ad.Tape(loss)
            boundary_condition()
            collision()
            streaming()
            compute_loss()
    if loss[None] == 0:
        break 
    # Gradient descent   
    inlet_pressure[None] += inlet_pressure.grad[None] * 
    learning_rate

\end{lstlisting}

 The gradients are propagated and evaluated through all variables and operations in the LBM simulation. Our simulations involve $10^4~10^7$ cells to model fluid flow through the granular media. We track all global tensors allocated to store the entire simulation state, including loss and inlet pressure scalars and 2D/3D tensors (density, velocity, force, pressure, \etc). Using `ti.root.lazy\_grad()', the velocity field (`v'), loss, inlet pressure, and other tensors involved in a derivative chain are automatically positioned in the adjoint fields in accordance with the layout of their primal fields

\section{Result and Discussion}

\subsection{Inferring boundary conditions from flow characteristics}
\subsubsection{Single-parameter inverse of boundary conditions}
\paragraph{Optimizing boundary conditions for pipe flow}

We demonstrate the capability of differentiable LBM to infer boundary conditions from observed flow velocities. We first simulate 2D laminar flow in a pipe with a thin obstacle plate as a test case. The aim is to estimate the constant pressure inlet boundary condition that drives this flow, using the resulting steady-state velocity field as a single-parameter inverse problem. 

\begin{figure}[htbp]
  \begin{subfigure}{\columnwidth}
    \centering
    \hspace*{-0.9cm}
    \adjincludegraphics[width=0.85\columnwidth,trim={0 0 0 0},clip]{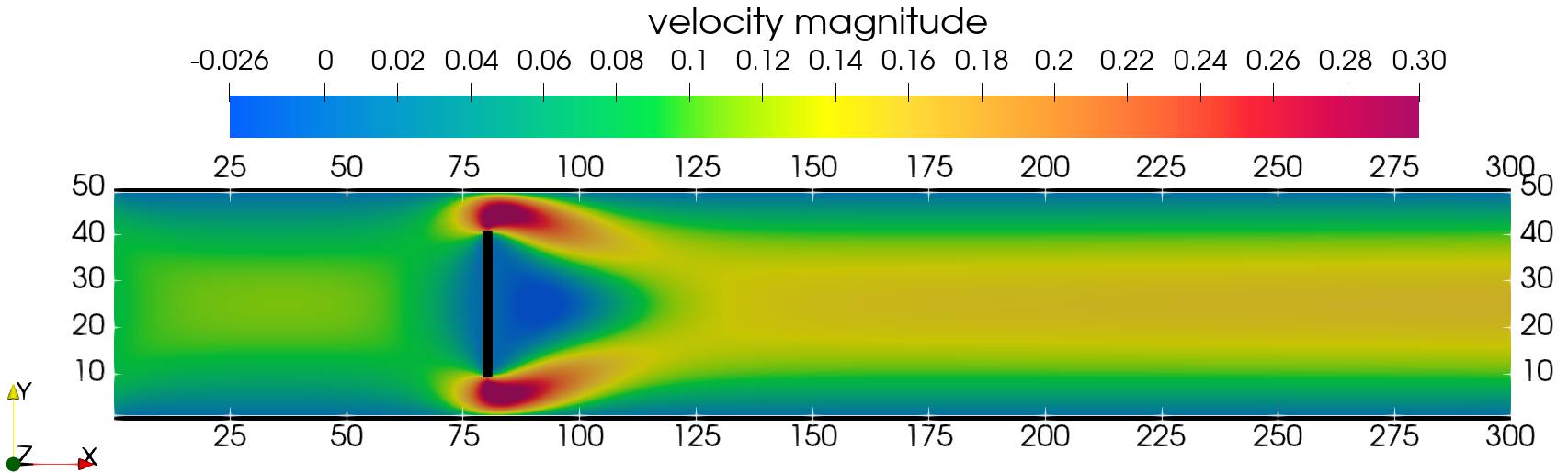}
    \caption{}
     \centering
     \adjincludegraphics[width=0.9\columnwidth,trim={0 0 0 {0.089\width}},clip]{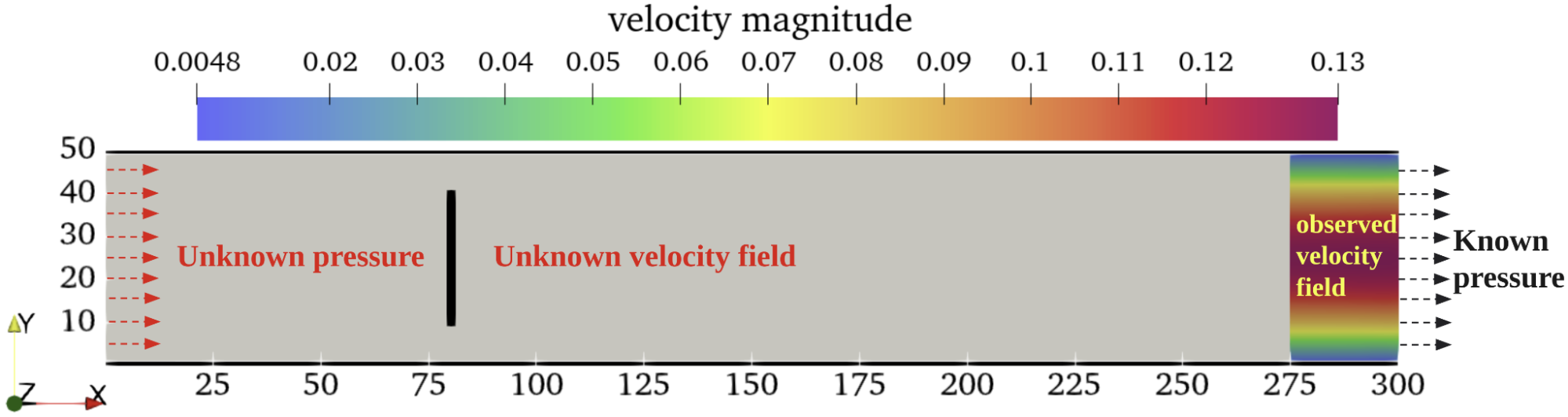}
     \caption{}
  \end{subfigure}%
\caption{AD-LBM simulation of steady laminar flow through a 2D pipe (width 50 lu, length 300 lu) containing an obstruction plate (width 1 lu, length  30 lu) situated at \( x = 80 \) lu: (a) velocity field driven by imposed inlet-outlet pressure difference of (mu)·(lu)$^{-1}$(ts)$^{-2}$; (b) unknown variables (inflow pressure and velocity field at \( x < 275 \) lu) and known variables (outflow  pressure and target outlet velocities field at \( x \geq 275 \) lu).}
\label{Fig:Poiseuille_flow_plate}
\end{figure}

\Cref{Fig:Poiseuille_flow_plate}a shows the steady-state velocity field of laminar flow in a 2D pipe (width 50 lu, length 300 lu) with a thin obstacle plate (width 1 lu, length 30 lu) at x = 80 lu. The flow is driven by a constant pressure difference of 0.05 mass units (mu)·length units (lu)$^{-1}$·time units (ts)$^{-2}$ between the inflow and outflow. We allow the velocity field 40,000 timesteps to reach steady state when the kinetic energy plateauing below 10$^{-4}$ (lu)(ts)$^{-1}$ standard deviation. The number of time steps required for the LBM to reach steady state is dependent on the spatial discretization of the fluid domain. A larger number of fluid nodes in the computational grid leads to an increased number of time steps needed to converge to a stable solution. 

To estimate the inlet pressure boundary, we initialize with a lower pressure of 0.0025 (mu)·(lu)$^{-1}$(ts)$^{-2}$ and compute the loss function between the simulated velocities and the target field. The target field is only observed at the outflow region (x $\geq$ 275 lu) to simulate limited sensor data, as shown in Figure 5b. Accordingly, we define the loss function by computing the difference between LBM velocities $u_{x,y}(p)$ and target velocities $|u_{x,y}^{target)}|$. It is defined as the summed squared errors over the lattice in this region of interest:
\begin{equation}
    loss(p)=\sum_{x}^{N_x}\sum_{y}^{N_y}\left [f(x,y)\cdot\left |u_{x,y}(p)  \right |-\left |u_{x,y}^{target}  \right |  \right ]^2,\,
\end{equation}
with $N_x$ and $N_y$ being the number of LBM grids in x and y dimensions, $|u_{x,y}(p)|$ is the norm of simulated velocity when the pressure p is applied at a specified boundary.  $|u_{x,y}^{target}|$ is the norm of target velocity. The function $f(x,y)$ defines the region of interest. For example, $f(x, y)$ is set to 1 for x $\geq$ 275 lu and 0 for x < 275 lu to only compute loss at the outlet region based on observed target data.

\begin{figure}
    \centering
    \includegraphics[width=0.7\columnwidth]{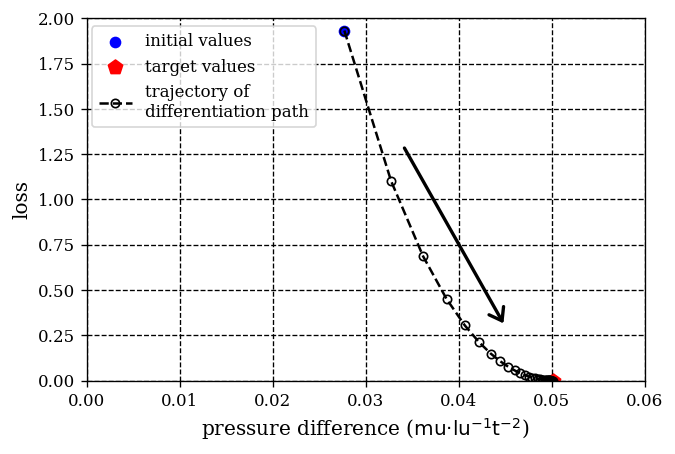}
    \caption{Directed Acyclic Graph (DAG) Representation of lattice Boltzmann Method Simulation Stages with two time steps and 2×2×2 domain division.}
    \label{Fig:6}
\end{figure}

Using the gradients from AD, we iteratively update the inlet pressure to minimize this loss function and match the observed target velocities.~\Cref{Fig:6} shows the optimization of inlet pressure boundary conditions using AD-LBM for obstructed pipe flow. The inlet pressure is updated each iteration to minimize loss between simulated and target outlet velocity fields, converging to the true pressure difference of 0.05 (mu)·(lu)$^{-1}$(ts)$^{-2}$ that generates the target flow. After 20 iterations using a learning rate of 0.05, the loss decreases to 10$^{-5}$ (lu)$^2$(ts)$^{-2}$. It demonstrates that even when using a limited target velocity field observation near the outlet, the AD-LBM approach efficiently determines the target boundary condition of 0.0025 (mu)·(lu)$^{-1}$(ts)$^{-2}$. 

\paragraph{Optimizing boundary conditions for for fluid flow through a granular media}

\begin{figure}[htbp]
  \centering
  \begin{subfigure}{0.3\columnwidth}
    \centering
    \includegraphics[width=\columnwidth]{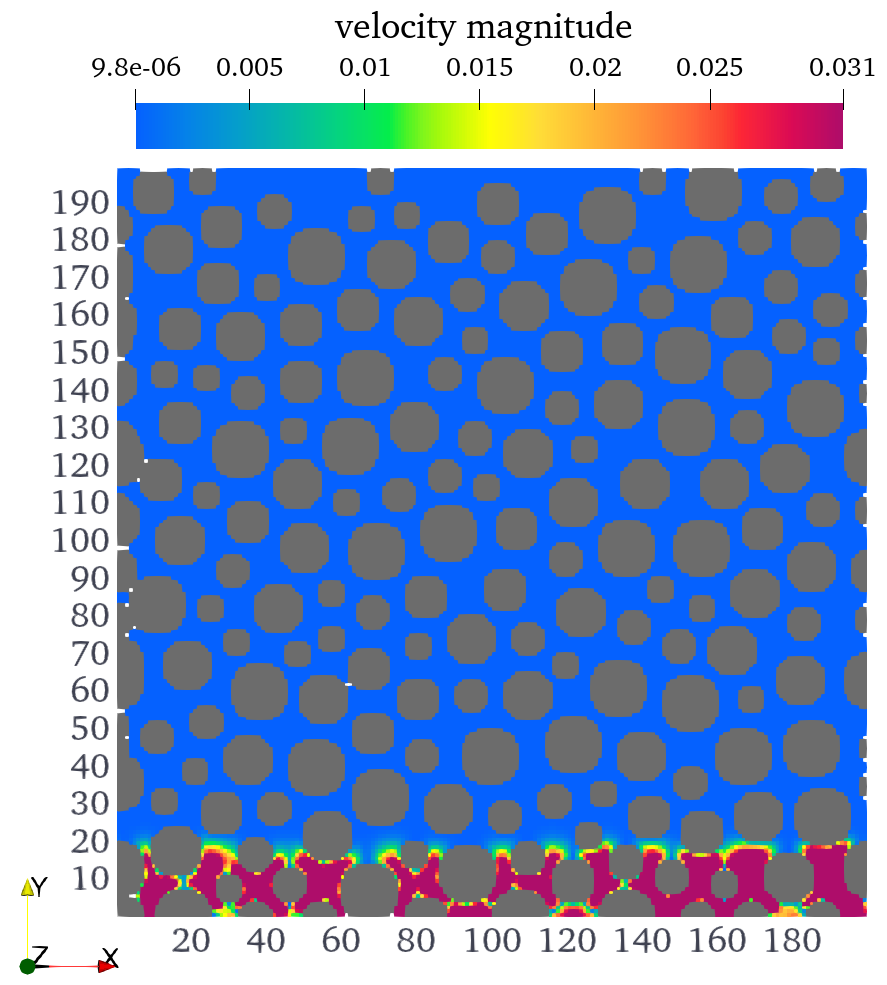}
    \caption{t = 0 ts}
  \end{subfigure}%
  ~
  \begin{subfigure}{0.3\columnwidth}
    \centering
    \adjincludegraphics[width=\columnwidth]{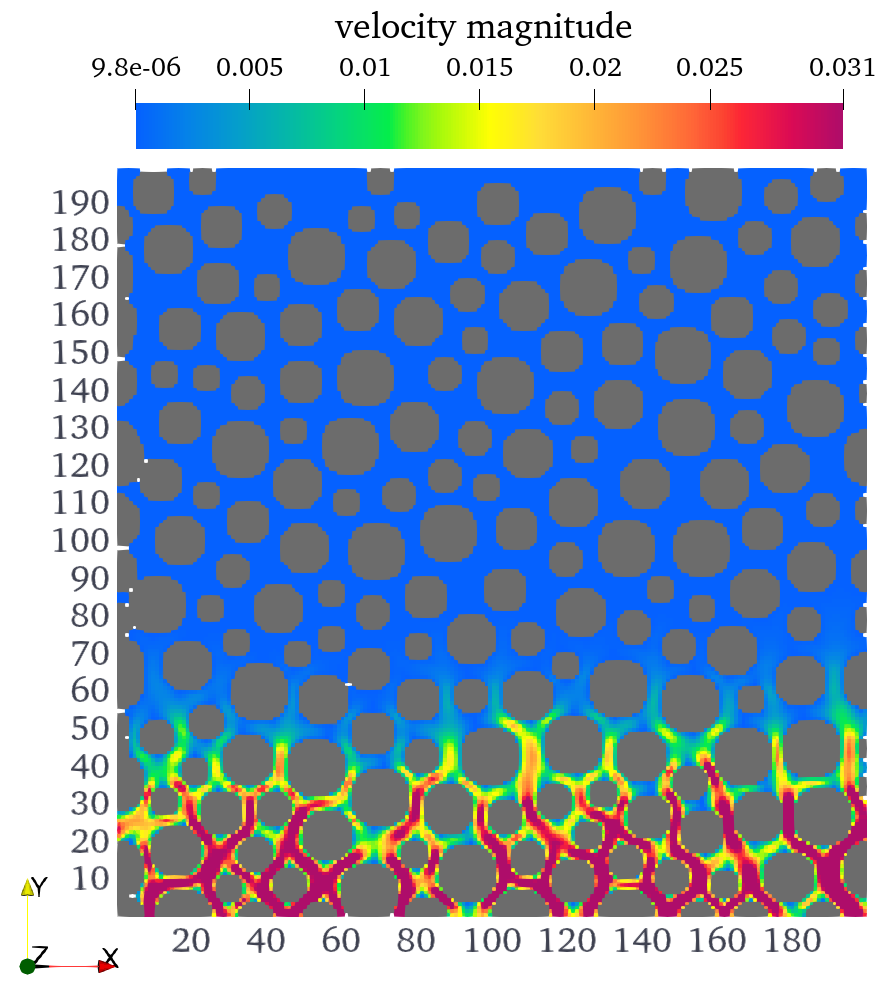}
    \caption{t = 100 ts}
  \end{subfigure}%
~
  \begin{subfigure}{0.3\columnwidth}
    \centering
    \includegraphics[width=\columnwidth]{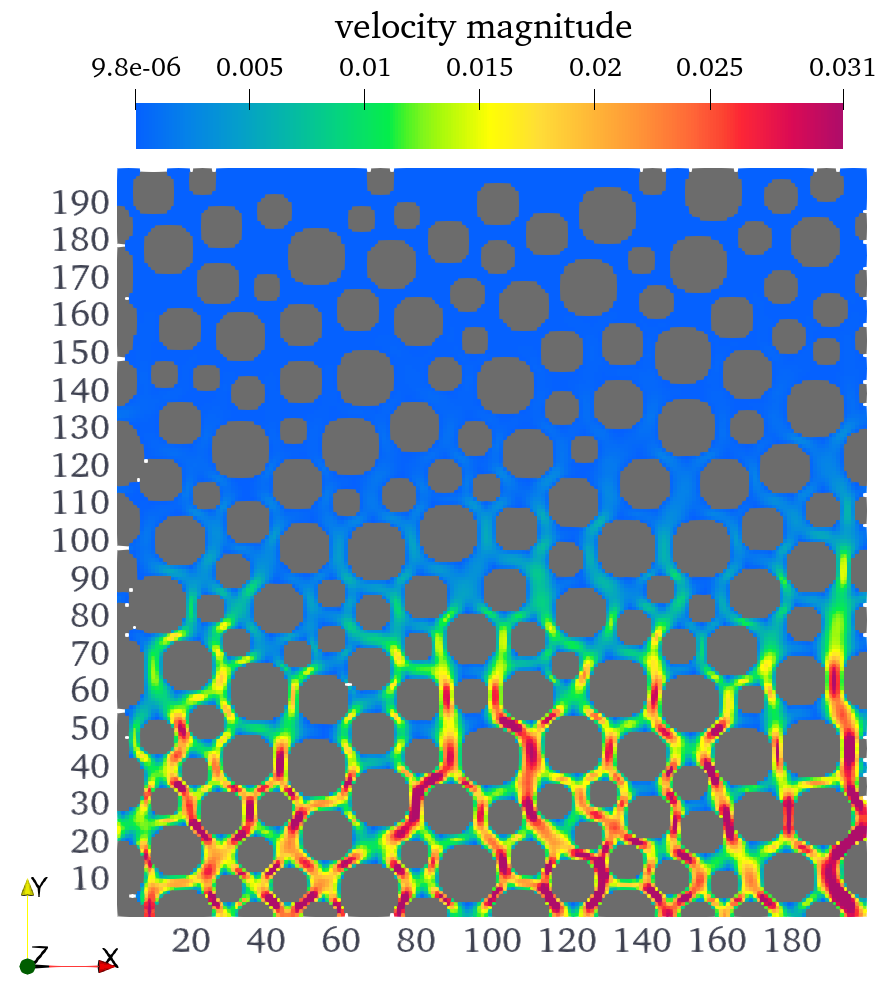}
    \caption{t = 500 ts}
  \end{subfigure}%

\ContinuedFloat
  \begin{subfigure}{0.3\columnwidth}
    \centering
    \adjincludegraphics[width=\columnwidth]{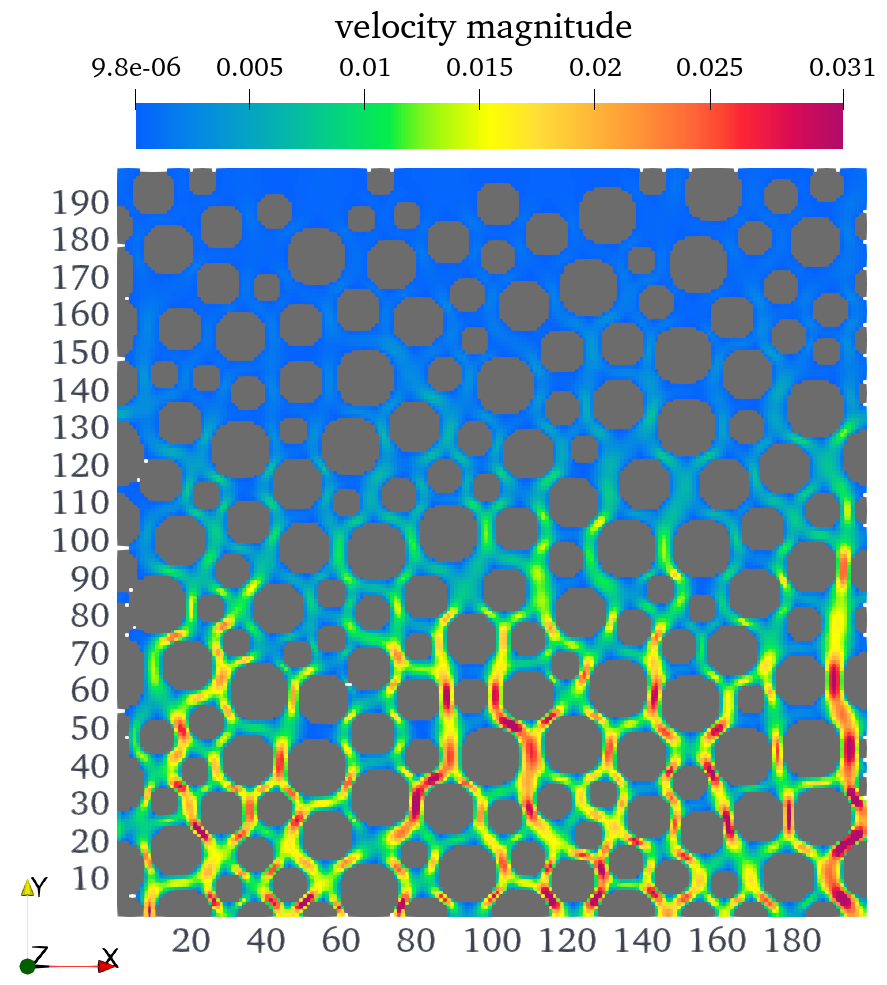}
    \caption{t = 1000 ts}
  \end{subfigure}
  ~
    \begin{subfigure}{0.3\columnwidth}
    \centering
    \includegraphics[width=\columnwidth]{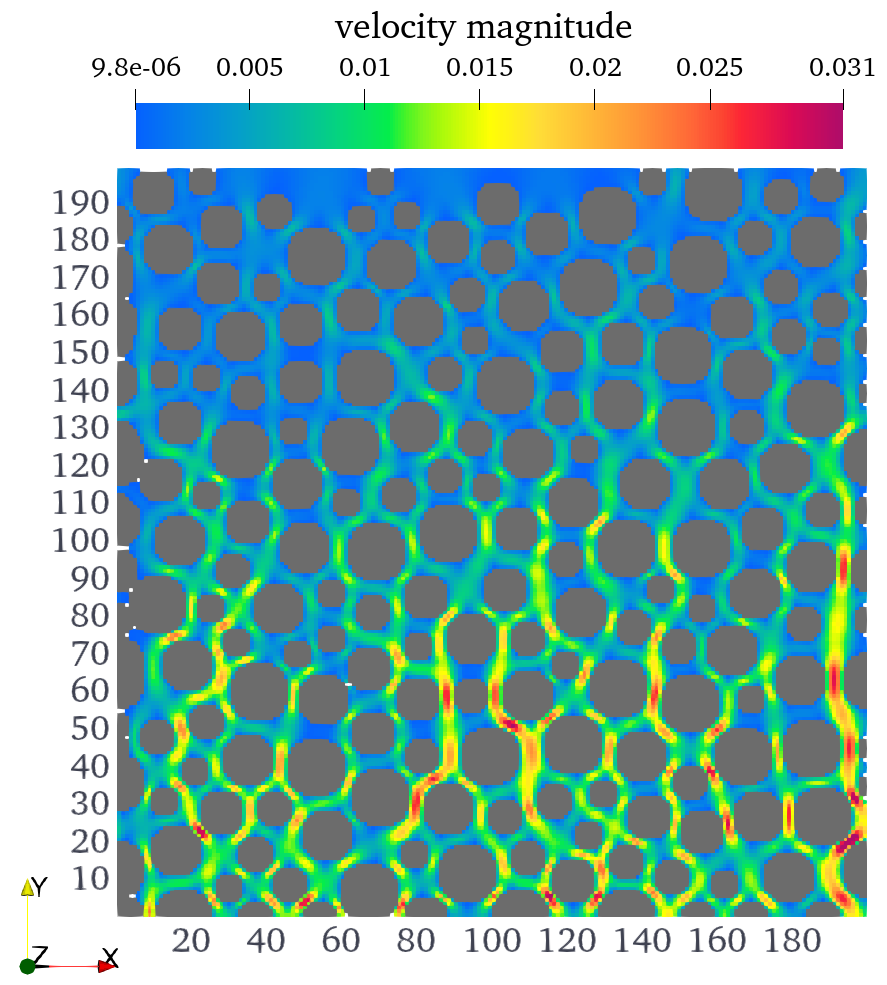}
    \caption{t = 2500 ts}
  \end{subfigure}%
  ~ 
  \begin{subfigure}{0.3\columnwidth}
    \centering
    \adjincludegraphics[width=\columnwidth]{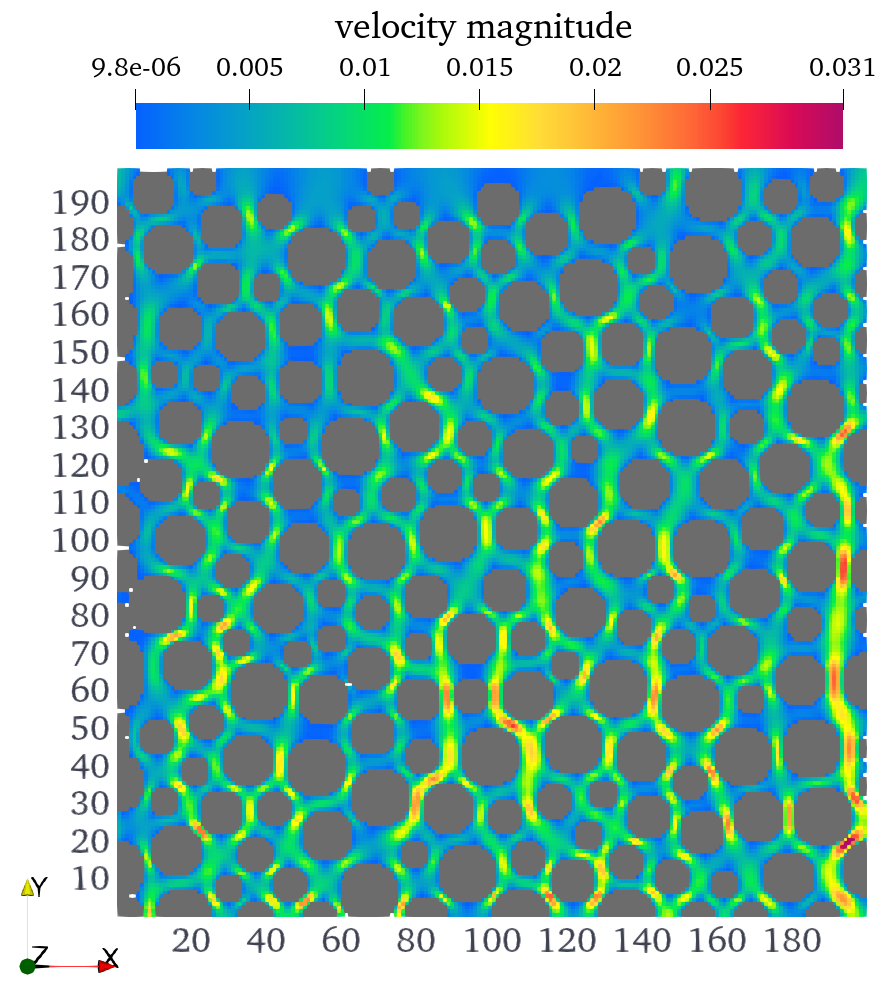}
    \caption{t = 5000 ts}
  \end{subfigure}
    
  \caption{Evolution of velocity field in LBM simulation for a laminar flow driven by a constant inlet pressure of (mu)·(lu)$^{-1}$(ts)$^{-2}$ through 2d granular packing.}
\label{Fig:2D_flow_evol}
\end{figure}

We now evaluate a more complex inverse problem of inferring the boundary condition for a laminar flow through a 2D granular packing using AD-LBM. When simulating a porous media flow in 2D, an important consideration is non-interconnected pore space when circular grains come in contact. In reality (3D), the pore spaces between the grains will allow fluid flow perpendicular to the reference plane. We solve the 2D artifact by adopting a hydrodynamic radius, a reduced radius of 0.9R only for the LBM computations~\citep{soundararajan2015multi}, where R is the actual radius of each grain. The hydrodynamic radius can be considered a 2D representation of the 3D permeability of the granular material. \Cref{Fig:2D_flow_evol} shows the evolution of the velocity field of a laminar flow driven by a constant pressure difference through 2D granular packing with the grains set immobile. This approach of immobilizing the grains enables isolating the effects of fluid flow on the granular media and a more focused investigation of the fluid flow dynamics. The flow is fully developed when the velocity reaches the top of the granular media. As seen in~\cref{Fig:2D_flow_evol}, the fluid flow reaches equilibrium after 5000 time steps.

\begin{figure}[htbp]
  \centering
  \begin{subfigure}{0.48\columnwidth}
    \centering
    \includegraphics[width=\columnwidth]{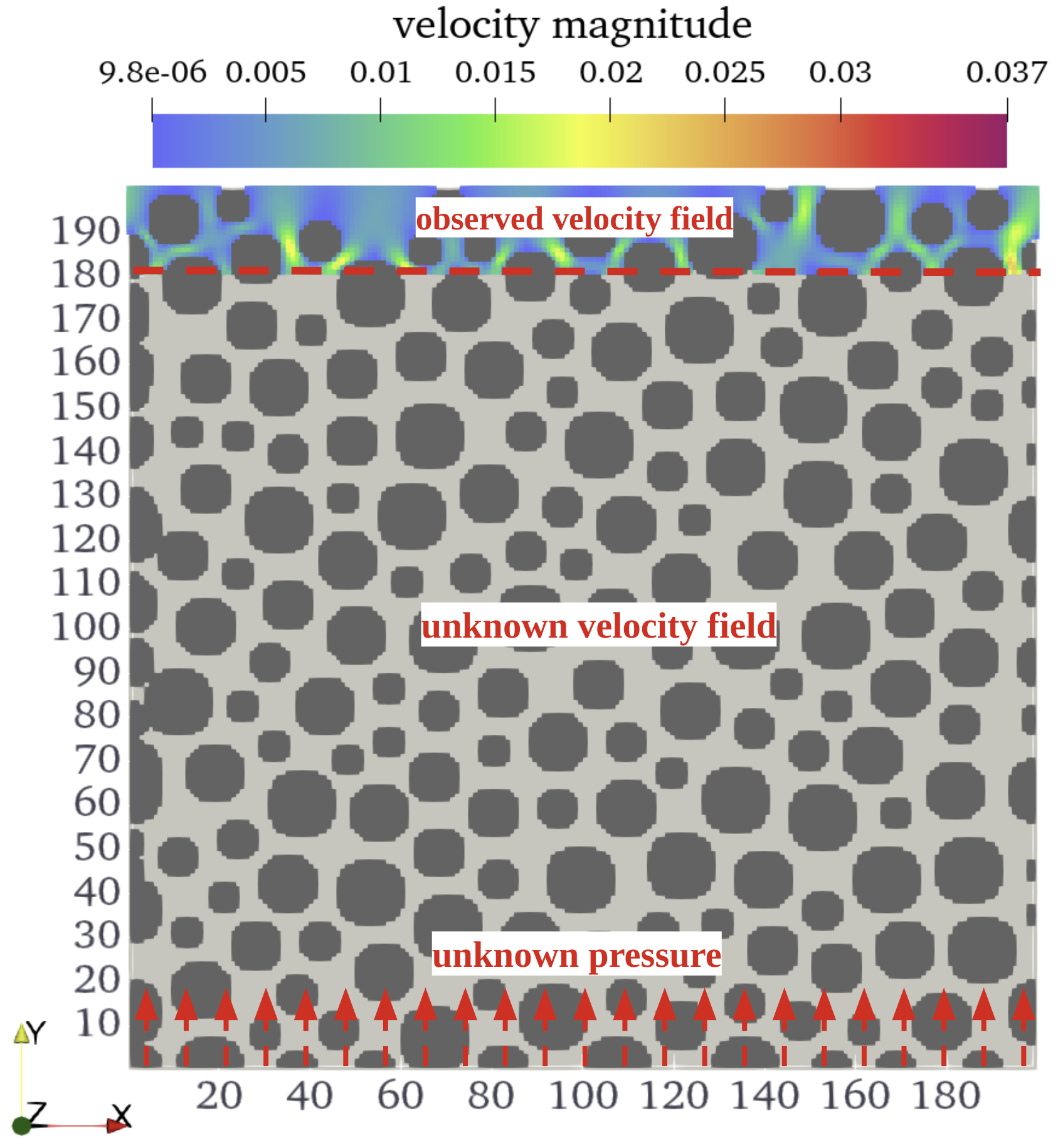}
    \caption{}
  \end{subfigure}%
  ~
  \begin{subfigure}{0.48\columnwidth}
    \centering
    \adjincludegraphics[width=\columnwidth]{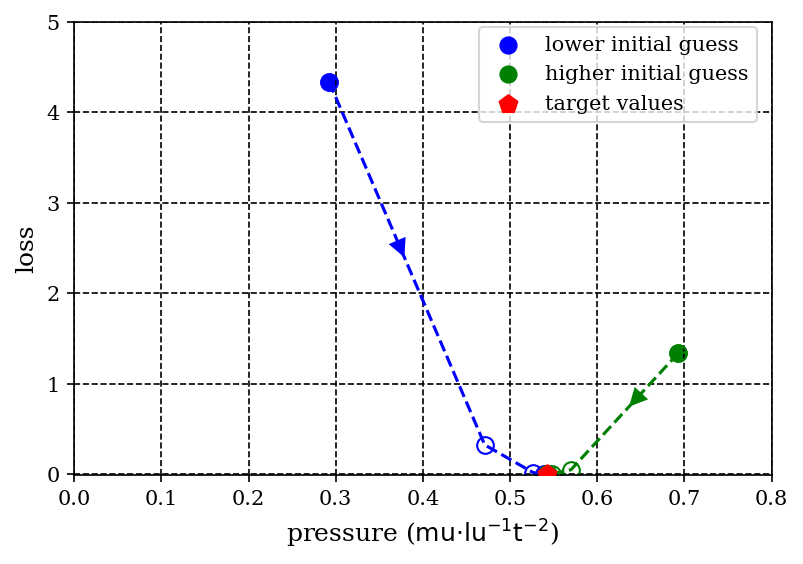}
    \caption{}
  \end{subfigure}%
  \caption{AD-LBM optimization of inlet pressure boundary conditions for laminar flow through 2d granular packing: (a) estimating unknown inlet pressure based on target outlet velocities (y $\geq $ 180 lu); (b) differentiation paths using two different inlet pressures.}
\label{Fig:8}
\end{figure}

We limit the target velocity field data to y $\geq $ 180 lu at time step 5000 ts, which yields velocity at 2300 fluid nodes.~\Cref{Fig:8} shows the evolution of velocity loss utilizing AD-LBM simulation. As illustrated in Figure 8a, the inlet pressure boundary condition was unknown, while the target velocity field and the granular packing were known. We initialized simulations using two different inlet pressures - one higher at 0.693 (mu)·(lu)$^{-1}$(ts)$^{-2}$ and one lower at 0.292 (mu)·(lu)$^{-1}$(ts)$^{-2}$ than the real target value. For inferring constant inlet pressure, the loss function decreased to 10-6 (lu)$^2$(ts)$^{-2}$ after 10 iterations, approaching the target pressure difference of 0.05 (mu)·(lu)$^{-1}$(ts)$^{-2}$, independent of initial guesses above or below the target as seen in Figure 8b. Therefore, the case studies of the obstructed pipe flow and flow through 2D granular packing demonstrate the ability of the AD-LBM to converge to an optimal solution for single-parameter optimization problems.

\subsubsection{Multi-parameter inverse of boundary conditions}

We now demonstrate a multi-parameter inversion using AD-LBM to identify a variable pressure boundary condition based on the target flow velocities at the top of a granular media. We apply a target pressure boundary as a cubic spline by varying the fluid densities at the inlet, as shown in Figure 9a. Our goal is to find the derivative of loss of the velocity profiles at the outlet with respect to the input pressure profile, given the observation of the target velocity field for y $>$ 180 lu. The 'inlet\_pressure' is no longer a constant but a 1D tensor of 200 elements, for x ranging from 0 lu to 199 lu and y = 0. 

\begin{figure}[htbp]
  \centering
  \includegraphics[width=0.5\columnwidth]{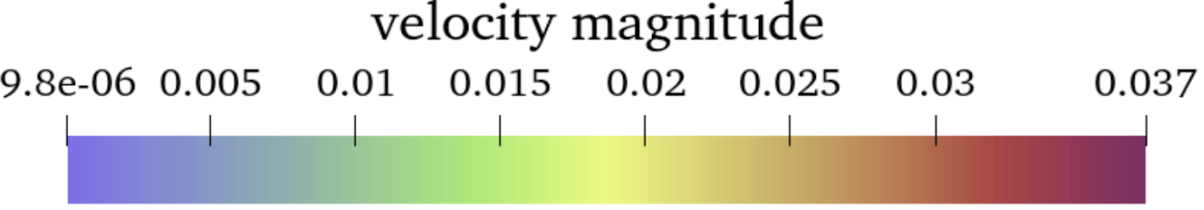}
  \begin{subfigure}{0.45\columnwidth}
    \centering
    \includegraphics[width=\columnwidth]{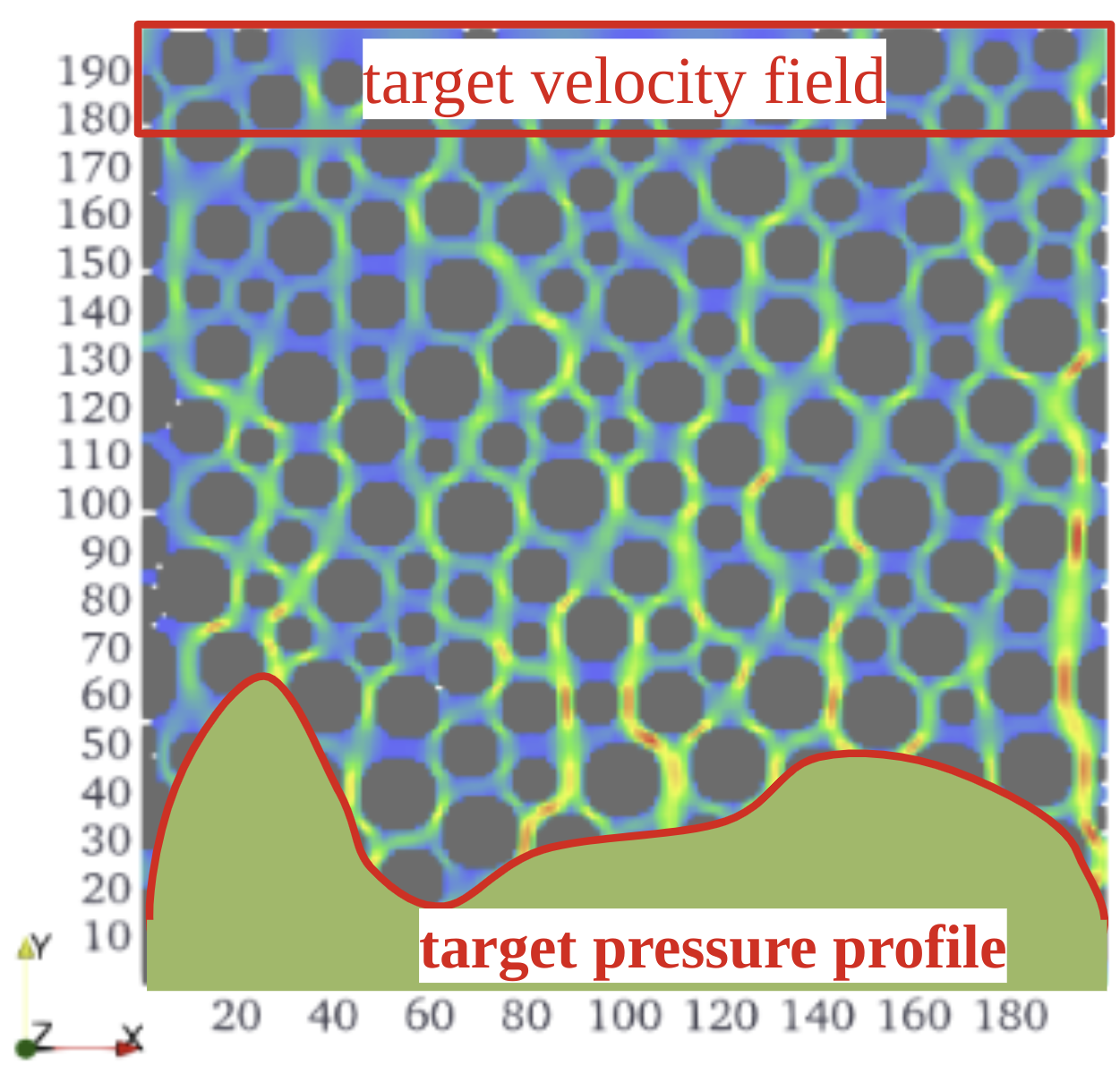}
    \caption{}
  \end{subfigure}%
  ~
  \begin{subfigure}{0.45\columnwidth}
    \centering
    \adjincludegraphics[width=\columnwidth]{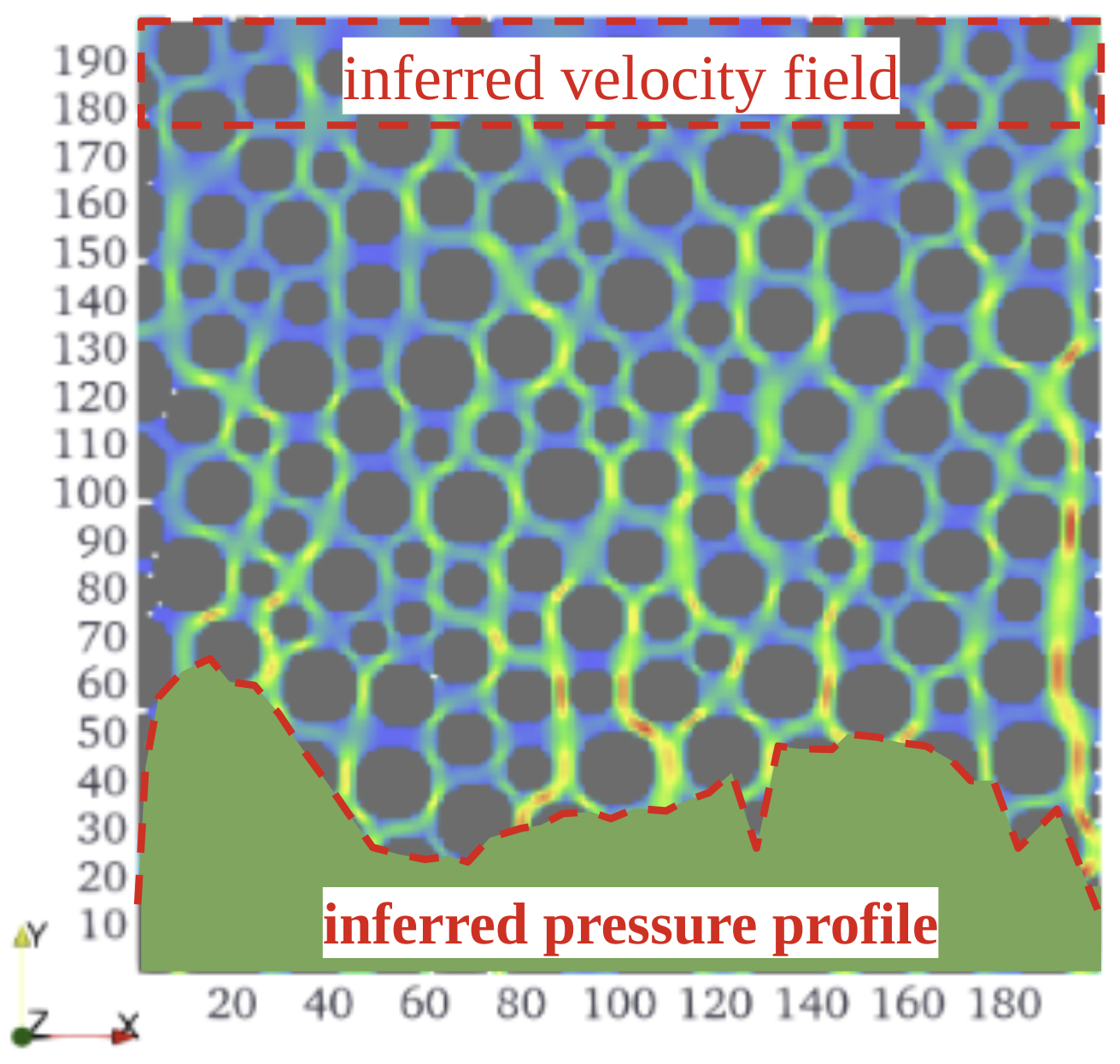}
    \caption{}
  \end{subfigure}%

  \begin{subfigure}{\columnwidth}
    \centering
    \includegraphics[width=0.75\columnwidth]{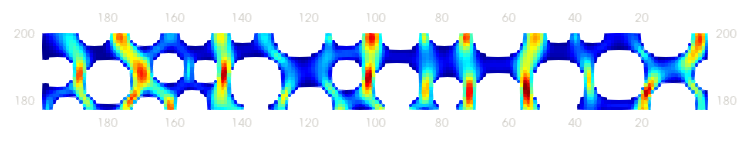}
    \includegraphics[width=0.75\columnwidth]{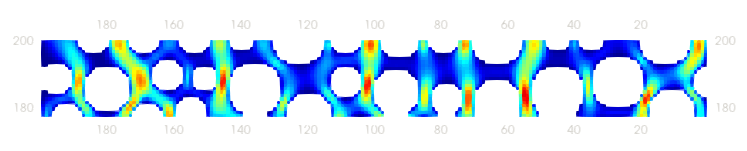}
    \caption{}
  \end{subfigure}
\end{figure}

\begin{figure}[H]
\ContinuedFloat
   \centering
    \begin{subfigure}{0.7\columnwidth}
    \includegraphics[width=\columnwidth]{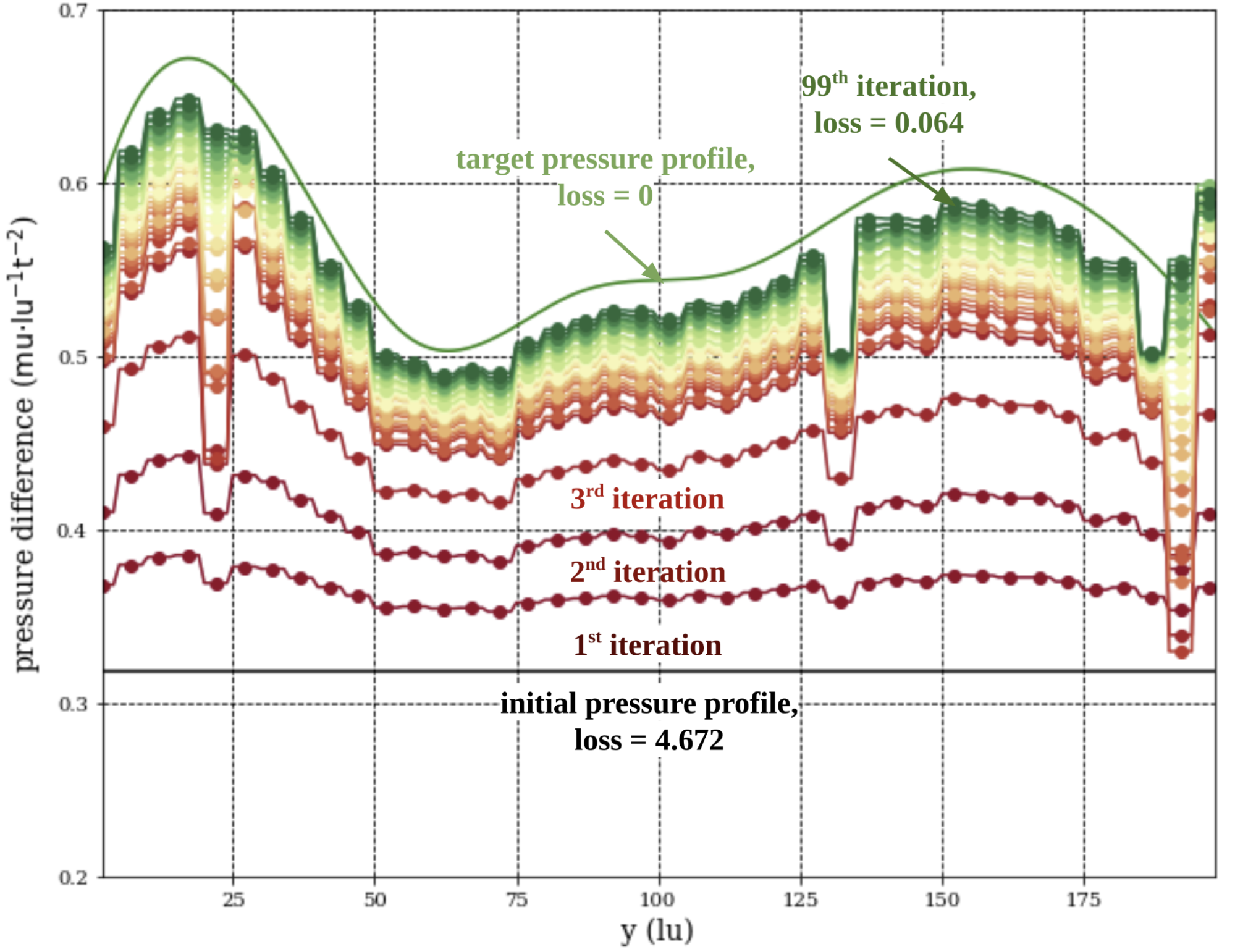}
    \caption{}
  \end{subfigure}%
  
  \caption{AD-LBM optimization of variable inlet pressure profile for 2D laminar flow through granular media based on outlet target velocity data (y $>$ 180 lu): (a) target pressure profile and induced velocity field; (b) inferred pressure profile and induced velocity field after 99 iterations; (c) comparison of the target velocity field (top) and inferred velocity field (bottom); and (d) loss and inferred pressure profiles for the first 100 iterations.}
\label{Fig:Pressure_profile}
\end{figure}

\Cref{Fig:Pressure_profile} illustrates AD-LBM optimization of a non-uniform inlet pressure profile based on limited target velocity data. Figure 9a-b shows the comparison of the target and inferred velocity field as well as the inlet pressure profile. Figure 9c compares the target velocity field (top) and the inferred velocity field after 99 iterations (bottom). Figure 9d shows the loss and inferred pressure profiles for the first 100 iterations. As an initial guess, we set the inlet pressure to be uniform at 0.32 (mu)·(lu)$^{-1}$(ts)$^{-2}$. We use a learning rate of 0.01 for optimization. The loss function requires 100 iterations to reduce from 4.62 (lu)$^2$(ts)$^{-2}$ to 0.064 (lu)$^2$(ts)$^{-2}$. After 100 iterations, the inferred inlet pressure profile matches the target pressure profile reasonably well, giving a root-mean-squared-error (RMSE) of 0.0254 (mu)·(lu)$^{-1}$(ts)$^{-2}$, $\text{RMSE} = \sqrt{\frac{1}{N} \sum_{i=1}^{N} (P_{\text{target},i} - P_{\text{inferred},i})^2}$ where N = 200 is the number of nodes at the inlet, Ptarget,i and Pestimated, i are the target and inferred pressure at x = i and y = 0, respectively.~\Cref{Fig:Pressure_profile}c-d also shows that the inferred velocity field matches the observed velocity field for the pressure profile with a velocity magnitude error of 0.064 (lu)$^2$(ts)$^{-2}$. Unlike the single-parameter cases, the loss function for the variable inlet pressure profile did not converge below $10^{-5}$ (lu)$^2$(ts)$^{-2}$, highlighting difficulties in matching non-uniform boundaries. Nevertheless, the consistency in the velocity field suggests that AD-LBM has promising capabilities in solving inverse problems for non-trivial boundary conditions in realistic flow scenarios. We can improve the slower convergence using a better initial guess through a neural network in future studies. 

\subsection{Approximation of Permeability for Porous Media}

In addition to inferring boundary conditions, the AD-LBM provides an integrated way to estimate the permeability of porous media. Each data point along the differentiation path represents a fully developed steady-state laminar flow solution. This enables permeability to be derived by directly applying Darcy's law to the AD-LBM simulations. We compared LBM-derived permeabilities against empirical solutions for 2D and 3D granular packings using the Carman-Kozeny (CK) equation, based on Poiseuille flow in a pipe:
\begin{equation}
k_{CK}=\Phi _{s}^{2}\frac{\varepsilon  ^{3}D_{p}^{2}}{\Phi_{CK}(1-\varepsilon )^2},\,
\end{equation}

where \( k_{CK} \) is the permeability, \( \varepsilon \) is the porosity, \( D_p \) is the mean grain diameter, \( \Phi_{CK} = 180 \) is the Kozeny shape factor~\citep{yazdchi2011microstructural}, and \( \Phi_{s} \) is the sphericity of the particles. \( \Phi_{s} \) equals 1 for purely spherical grains. For the 2D granular packing shown in~\cref{Fig:Pressure_profile}, \( D_p = 1.18 \) mm and \( \varepsilon = 0.489 \). Substituting these values into Eq. (8) gives an analytical permeability of \( k_{CK} = 3.164 \times 10^{-9} \) m\(^2\). Note that the conversion from lattice units to physical units is \( \frac{L_{\text{physical}}}{L_{\text{LBM}}} = 10^{-4} \) m·(lu)\(^{-1}\), where \( L_{\text{physical}} \) is the physical length in meters, and \( L_{\text{LBM}} \) is the lattice length in lattice units.

Converting the LBM-derived permeability to physical permeability requires defining consistent LBM units for mass (\text{mu}), time (\text{ts}), and density to match physical fluid properties. The fluid domain is initialized with a uniform density of \( 1 \, \text{mu} \cdot \text{(lu)}^{-3} \), corresponding to the density of water (\( 1000 \, \text{kg} \cdot \text{m}^{-3} \)). The mean fluid density \( \rho_{\text{fluid}} \) represents the total fluid mass divided by the volume and depends on the pressure difference \( \partial P \). A higher \( \partial P \) increases \( \rho_{\text{fluid}} \). With \( \partial P = 0 \), the volume of \( 1 \, \text{(lu)}^3 \) has the mass \( \rho_{\text{fluid}} \times 1 \, \text{(lu)}^3 = 1 \, \text{mu} \) which equals \( 1000 \, \text{kg} \cdot \text{m}^{-3} \times (10^{-4} \, \text{m} \cdot \text{(lu)}^{-1})^3 = 10^{-9} \, \text{kg} \). Thus, \( 1 \, \text{mu} = 10^{-9} \, \text{kg} \) for \( \partial P = 0 \). For a \( \partial P = 0.276 \, \text{mu} \cdot \text{(lu)}^{-1} \cdot \text{(ts)}^{-2} \), \( \rho_{\text{fluid}} = 1.40 \, \text{mu} \cdot \text{(lu)}^{-3} \). This gives \( 1.40 \, \text{mu} \cdot \text{(lu)}^{-3} \times 1 \, \text{(lu)}^3 = 10^{-9} \, \text{kg} \), so \( 1 \, \text{mu} = 7.14 \times 10^{-10} \, \text{kg} \).

We use the correlation between the kinematic viscosity of a fluid to determine the time in the physical and lattice scales. The kinematic viscosity of water equals and $\nu_{physical}$= 10$^{-6}$ m$^2$/s in physical units. In contrast, the kinematic viscosity of water in LBM is defined as
\begin{equation}
\nu_{LBM}=c_{s}^{2}(\tau - \frac{\Delta t}{2}),\,
\end{equation}
where \( \tau \) is the dimensionless relaxation time and \( c_s \) is the isothermal speed of sound, given as \( c_s = \frac{1}{3} \frac{\Delta x}{\Delta t} \) for the velocity sets used. We use the BGK operator, hence, \( \tau = \Delta t \). Therefore, \( \nu_{\text{LBM}} \) can be written as:
\begin{equation}
\nu_{LBM}=\frac{\Delta x^2}{6\Delta t},\,
\end{equation}

where \( \Delta x \) and \( \Delta t \) are set to unity for simplicity, giving \( \Delta x = 1 \, \text{lu} \) and \( \Delta t = 1 \, \text{ts} \). By setting \( \nu_{\text{LBM}} = \nu_{\text{physical}} \), \( 1 \, \text{ts} = 1.67 \times 10^{-3} \, \text{s} \). Thus, the pressure difference of \( \partial P = 0.276 \, \text{mu} \cdot \text{(lu)}^{-1} \cdot \text{(ts)}^{-2} \) corresponds to a pressure gradient of \( \frac{\partial P}{\partial x} = 1.38 \times 10^{-3} \, \text{mu} \cdot \text{(lu)}^{-2} \cdot \text{(ts)}^{-2} = 35.36 \, \text{Pa} \cdot \text{m}^{-1} \). The mean fluid velocity \( U = 3.88 \times 10^{-3} \, \text{mu} \cdot \text{(ts)}^{-1} = 2.33 \times 10^{-4} \, \text{m} \cdot \text{s}^{-1} \). With the knowledge of channel width \( D = 200 \, \text{lu} = 0.02 \, \text{m} \), mean fluid velocity \( U \), and fluid kinematic viscosity \( \nu \), we derive the dimensionless Reynolds number as:

\begin{equation}
\begin{split}
R_e = &\frac{UD}{2\nu}\\
= &\frac{3.88\times10^{-3}\,\text{mu/ts}\times200\,\text{lu}}{2\times0.167\,\text{mu}^2/\text{ts}}\\
=&\frac{2.33\times10^{-4}\,\text{m/s}\times0.02\,\text{m}}{2\times10^{-6}\,\text{m}^2/\text{s}}=2.33,\,
\end{split}
\end{equation}

Since $R_e$ is as low as 2.33, the flow simulated by LBM is laminar and comparable to the physical flow. We can calculate the permeability using Darcy's law:
\begin{equation}
    k_{LBM}=\frac{\rho_{fluid}\nu U}{\frac{\partial p}{\partial x}},\,
\end{equation}

Accordingly, the AD-LBM yields kLBM = 0.317 (mu)$^2$, corresponding to the physical permeability  $k_{physical} = 3.17 \times 10^{-9}\text{m}^2$. The permeability determined by LBM simulation is extremely close to the permeability predicted by the CK model ($k_{CK} = 3.164 \times 10^{-9}\text{m}^2$). The permeability value calculated from LBM simulation is in dimensionless lattice unit, and it may be converted to physical unit using~\citep{sukop2013evaluation}:
\begin{equation}
    k_{physical}=k_{LBM}(\frac{L_{physical}}{L_{LBM}})^2,\,
\end{equation}
By substituting $\frac{L_{physical}}{L_{LBM}}= 10^{-4}$ m/lu, we can easily derive the same value of $k_{physical}$. 


\begin{table}[htbp]
\centering
\caption{Unit conversion of variables for 2D laminar flow problem.}
\begin{tabular}{@{}>{\raggedright\arraybackslash}p{6cm}>{\raggedright\arraybackslash}p{3.5cm}>{\raggedright\arraybackslash}p{4.5cm}@{}}
\toprule
\multicolumn{3}{c}{\textbf{Reference Variables in Physical-Scale and Lattice-Scale}} \\
\midrule
Reference Variables & Physical-Scale & Lattice-Scale \\
\midrule
Mass & \(7.14 \times 10^{-10}\) kg & 1 mu \\
Time & 0.00167 s & 1 ts \\
Length & \(10^{-4}\) m & 1 lu \\
\midrule
\multicolumn{3}{c}{\textbf{Converting properties between scales}} \\
\midrule
Properties & Physical-Scale & Lattice-Scale \\
\midrule
Reynolds number (\(R_e\)) & 2.33 & 2.33 \\
Mean fluid density (\(\rho_{\text{fluid}}\)) & 1000 kg\(\cdot \text{m}^{-3}\) & 1.40 \( \text{(mu)} \cdot \text{(lu)}^{-3} \) \\
Pressure gradient (\(\partial P/\partial x\)) & 35.36 Pa\(\cdot \text{m}^{-1}\) & \(1.38 \times10^{-3} \text{(mu)}\cdot\text{(lu)}^{-2}\cdot\text{(ts)}^{-2} \) \\
Kinematic viscosity (\(\nu\)) & \(10^{-6} \text{m}^2\cdot \text{s}^{-1}\) & 0.167 \( \text{(lu)}^2\cdot\text{(ts)}^{-1} \) \\
Dynamic viscosity (\(\mu\)) & \(10^{-3} \text{Pa}\cdot \text{s}^{-1}\) & 0.234 \( \text{(mu)}\cdot\text{(lu)}^{-1}\cdot\text{(ts)}^{-1} \) \\
Mean fluid velocity (U) & \(2.33\times 10^{-4} \text{m}\cdot \text{s}^{-1}\) & \(3.88\times10^{-3} \text{(lu)}\cdot\text{(ts)}^{-1} \) \\
Permeability (k) & \(3.47\times10^{-9} \text{s}^2\) & 3.47 \( \text{(mu)}^2 \) \\
Hydraulic conductivity (K) & 3.17 \( \text{cm}\cdot \text{s}^{-1} \) & \(3.17 \times10^8 \text{(mu)}\cdot\text{(ts)}^{-1} \) \\
\bottomrule
\end{tabular}
\label{tab:Unit_convert}
\end{table}

We evaluate the permeability of the granular media by determining the pressure gradient variation at each iteration of AD-LBM. The target pressure gradient is \( 35.36 \, \text{Pa} \cdot \text{m}^{-1} \), and the pressure difference is \( 0.707 \, \text{Pa} \). At each iteration, we calculate mean fluid density and update the mass units in both the physical scale and lattice scales, while the length and time units remain fixed. We then convert the mean flow velocity and pressure gradient from lattice units to physical units (see~\cref{tab:Unit_convert}).~\cref{Fig:2D_permeability} shows the relationship between the loss function and the pressure difference, as well as the mean flow velocity and pressure gradient for each iteration. The initial estimation of pressure difference is \( 0.087 \, \text{Pa} \); after 20 iterations of differentiation, the loss reaches \( 10^{-4} \, \text{(lu)}^2 \cdot \text{(ts)}^{-2} \). According to Darcy's law, the permeability can be derived as a product of the kinematic viscosity, porosity, and the proportionality constant between the mean fluid velocity and the applied constant pressure gradient.

\begin{figure}[htbp]
  \begin{subfigure}{0.49\columnwidth}
    \centering
    \includegraphics[width=\columnwidth]{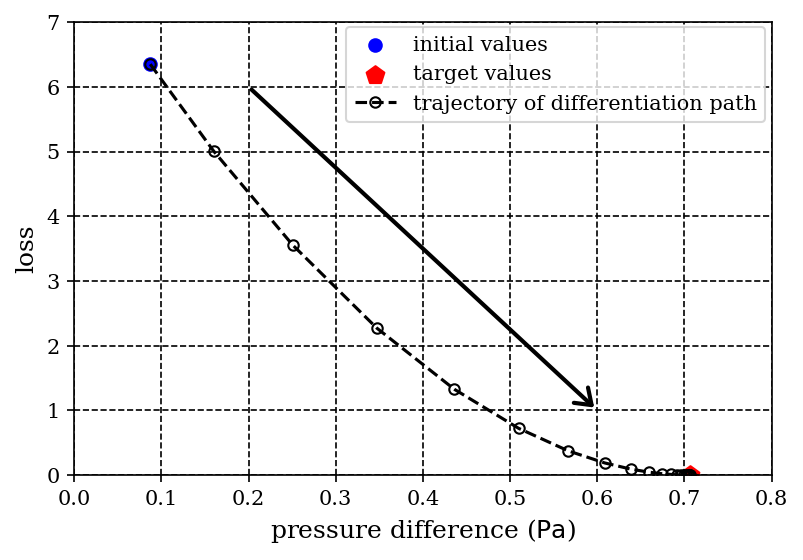}
    \caption{}
  \end{subfigure}%
  ~
  \begin{subfigure}{0.49\columnwidth}
    \centering
    \includegraphics[width=\columnwidth]{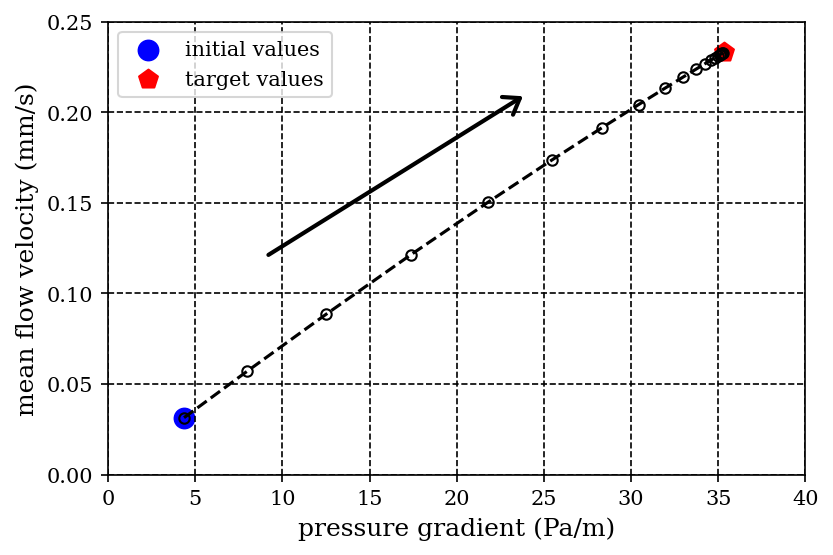}
    \caption{}
  \end{subfigure}%
\caption{AD-LBM analysis of laminar flow in 2D granular packaging: (a) differentiation path and (b) relationship between mean flow velocity and pressure gradient.}
\label{Fig:2D_permeability}
\end{figure}

\subsection{Inverse analysis in 3D granular media}

We extend the AD-LBM to 3D laminar flow through granular media for more realistic geometries. We demonstrate the capability of the 3D AD-LBM in evaluating pressure boundaries from limited flow velocities and estimating the permeability of 3D granular media of real Hamburg sand.  We develop a digital twin of the 3D sample using the technique described in~\citep{wang2022multiphase}. We use phase segmentation on the 12-mm-diameter cylinder sample to produce a digital twin of the CT experiment using the LBM. Figure 11a shows the grain structure. We store the segmented data in Tag Image File Format (TIFF), which comprises 1200 photos of the cross-section every 0.01 mm along the sample height. Each slice represents a labeled volume, with each voxel labeled with an integer value (0 for pore space and 1 for solid grain). We then select the representative elementary volume (REV) from the cylinder specimen. The LB REV model represents the 6-mm-high central cubical sub-volume. The grains have a mean diameter of 0.68 mm and a porosity of 0.36~\citep{wang2022multiphase}.

We employ the D3Q19 scheme for velocity discretization, which is more computationally intensive despite being GPU-capable. We restrict its size to 200 lu (= 6 mm) in height and 100 lu (= 3 mm) in width and length. Thus, $\frac{L_{\text{real}}}{L_{\text{LBM}}} = 3 \times 10^{-5} m\cdot lu^{-1}$. We apply a pressure gradient of $2.94 \times 10^{-3}$ mu $\cdot$ lu$^{-1} \cdot$ ts$^{-2}$, equivalent to $1.874 \text{kPa}\cdot m^{-1}$ in physical units, which creates a fully developed laminar flow with $R_e = 3.85$."

To estimate permeability using the CK model for the 3D granular packing, we determine the Kozeny factor $\Phi_{CK}$ based on the tortuosity:
\begin{equation}
    \Phi_{CK}=\Phi_ (\frac{L_e}{L})^2,\,
\end{equation}

where $\Phi$ accounts for particle shape effects ($\Phi = 90$ for spherical particles). The tortuosity $\frac{L_e}{L}$ depends on the streamline length $L_e$ relative to sample length L. Using ParaView on the stored VTK files, we extract streamlines traversing 200 lu pore space length along the REV sample. By integrating along each steady-state streamline as shown in Figure 11b, and averaging, we obtained a tortuosity of $\frac{L_e}{L}$  = 1.441. With $\Phi = 90$, this gives a Kozeny factor of $ \Phi_{CK}$ = 186.63. Substituting into the CK equation results in an estimated permeability of $k_{CK}=2.82 \times 10^{-10}$ m$^2$ for the 3D packing.

\begin{figure}[htbp]
  \begin{subfigure}{0.74\columnwidth}
    \includegraphics[width=\columnwidth]{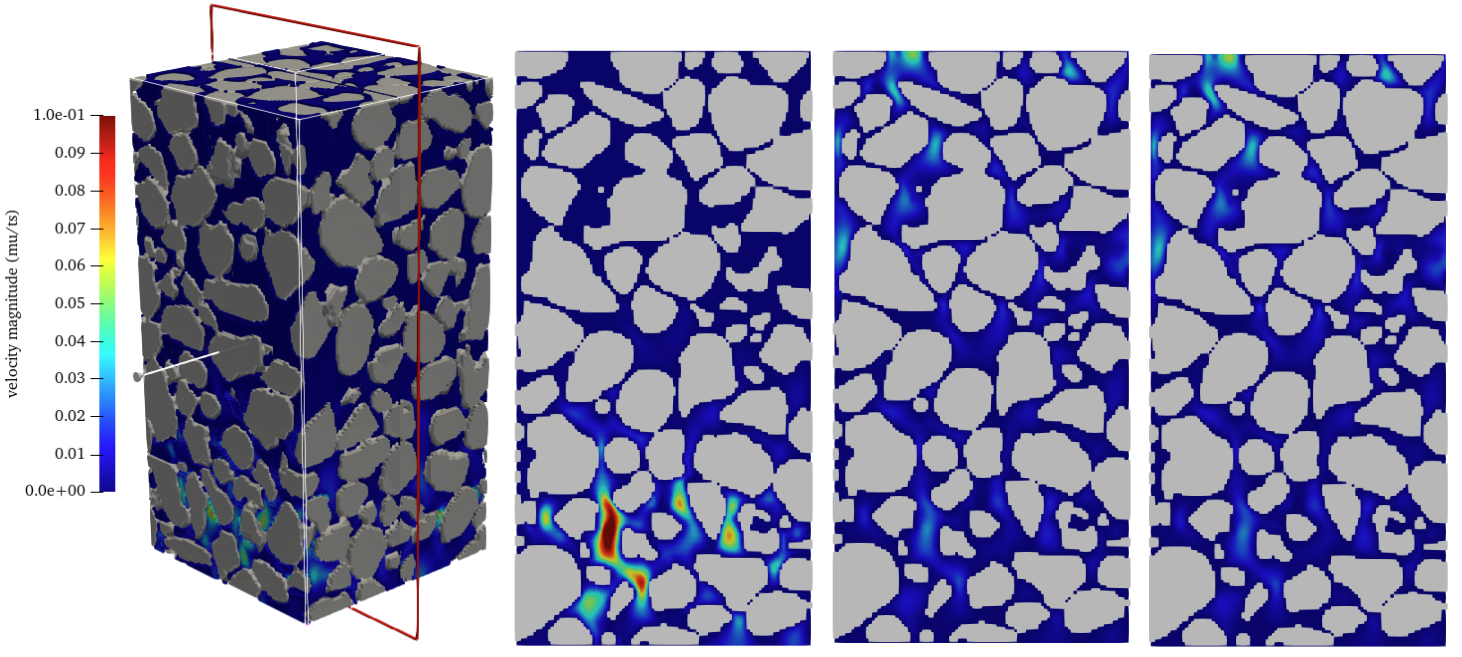}
    \caption{}
  \end{subfigure}%
  \begin{subfigure}{0.23\columnwidth}
    \includegraphics[width=\columnwidth]{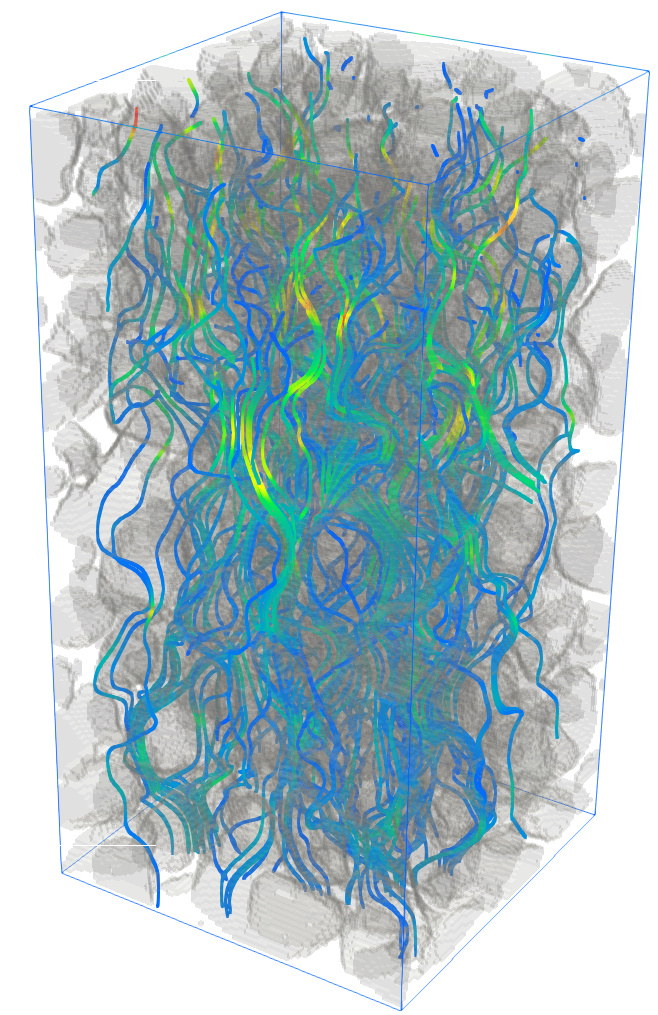}
    \caption{}
  \end{subfigure}%

  \begin{subfigure}{0.5\textwidth}
   \ContinuedFloat
    \centering
    \includegraphics[width=\linewidth]{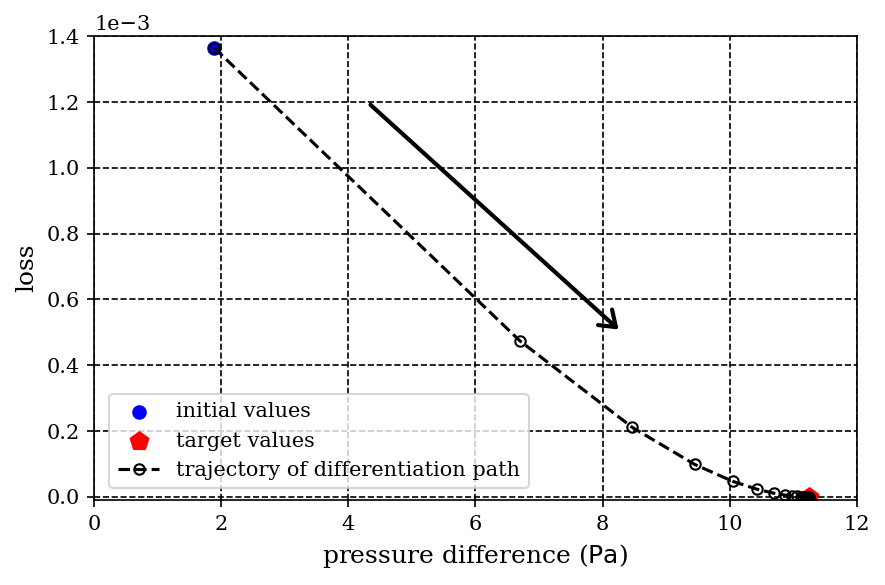}
    \caption{}
  \end{subfigure}%
  \begin{subfigure}{0.48\textwidth}
    \centering
    \includegraphics[width=\linewidth]{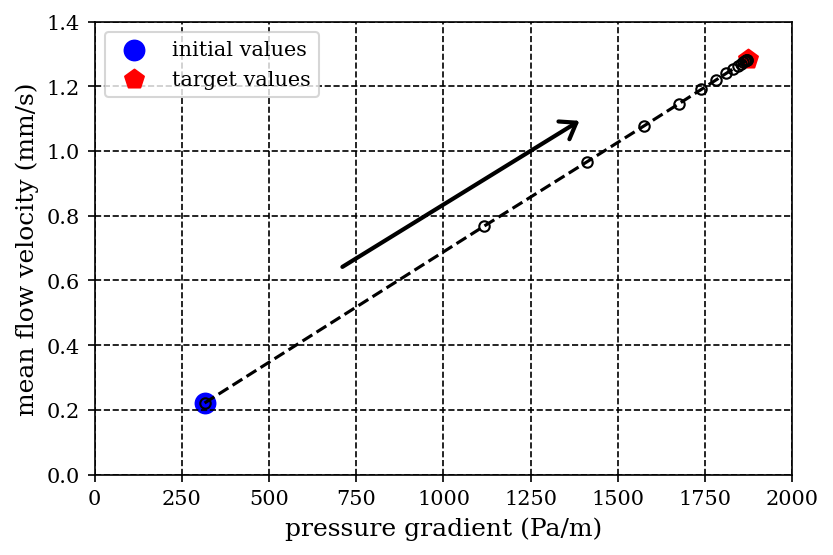}
    \caption{}
  \end{subfigure}%
\caption{AD-LBM simulation of laminar flow through the Hamburg sand: (a) development of the flow through the 3D granular media; (b) steady-state streamline; (c) path of differentiation trajectory of loss function; and (d) mean flow velocity versus pressure gradient.}
\label{Fig:3D_hamburg_flow}
\end{figure}

For the 3D case, the loss function comparing simulated and target velocities is:
\begin{equation}
 loss(p)=\frac{1}{N_{fluid}}\sum_{x}^{N_x}\sum_{y}^{N_y}\sum_{z}^{N_z}\left [\left |u_{x,y}(p)  \right |-\left |u_{x,y}^{target}  \right |  \right ]^2,\,
\end{equation}
where $N_{fluid}$ is the number of fluid nodes. We use normalization of loss to improve convergence for the larger domain size. The 3D LBM model includes 761,855 fluid nodes ($N_{fluid}$ = 761,855) compared to $N_{fluid}$ = 19,200 in 2D. Though normalization does not impact the optimized pressure, it provides better-scaled gradients for the iteration updates. 

\cref{Fig:3D_hamburg_flow}a illustrates the progressive permeation of the laminar flow through the 3D Hamburg sand's pore spaces over \(20,000\) LBM timesteps, due to the applied pressure gradient. The flow achieves an average outlet velocity of \(0.4 \, \text{lu} \cdot \text{(ts)}^{-1}\). The loss function decreased to \(10^{-6} \, \text{(lu)}^2 \cdot \text{(ts)}^{-2}\) after 20 iterations, yielding an estimated permeability of \(0.285 \, \text{mu}^2\) or \(2.56 \times 10^{-10} \, \text{m}^2\), within \(4\%\) of the CK analytical estimate of \(2.82 \times 10^{-10} \, \text{m}^2\). Overall, these 3D results further demonstrate the capability of AD-LBM in inferring boundaries for realistic intricate geometries.

Moreover, the GPU-accelerated differentiable fluid simulator demonstrates significant speedups over multi-threaded CPU lattice Boltzmann methods. In benchmark tests with \(5,000\) kernel invocations for \(761,855\) nodes, the total CPU execution time was \(234.4\) seconds using \(56\) threads. This was dominated by three main kernels -- streaming (\(41.6\%\)), collision (\(26.0\%\)) and boundary conditions (\(25.1\%\)). By leveraging a Quadro RTX 5000 GPU and CUDA parallelization, the total runtime was slashed to \(46.9\) seconds, a \(5\times\) speedup over CPU. The highly parallelized streaming kernel accounted for \(52.9\%\) of GPU execution time. Other top GPU kernels were boundary conditions (\(29.5\%\)) and collision (\(12.1\%\)). This prominent \(5\times\) simulation speedup highlights the benefits of massively parallel architectures for computational fluid dynamics. Our GPU-based differentiable simulator enables faster iteration on larger, more complex fluid systems previously hindered by computational cost.

\subsection{Material property determination}

In this section, we evaluate the fluid rheological properties, such as its viscosity, based on the flow characteristics and known boundary conditions. We will employ the AD-LBM with the Multiple Relaxation Time (MRT) operator to consider turbulent flows~\citep{lallemand2000theory}. In LBM-MRT, the advection is mapped onto the momentum space by a linear transformation, and the flux is finished within the velocity space~\citep{du2006multi}. The MRT collision operator, different from the BGK operator, is:
\begin{equation}
   \Omega _{col}=M^{-1}S(Mf-m^{eq}),\,
\end{equation}

The collision process is conducted in the momentum space, and the equilibrium moment is calculated by $m^{eq}=Mf^{eq}$. Alternatively, one can also construct equilibrium moments more precisely and efficiently from the known moments and u using a general polynomial representation for D2Q9 model:
\begin{equation}
   m_i^{eq}=\rho\sum_{l,m}a_{i,l,m}mu_x^lu_y^m,\,
\end{equation}
where $i=0,1,2,...,8$ denotes the discrete direction. The exact form of the equilibrium moments, referring to the coefficients $a_{i,l,m}$, are constructed via Gram-Schmidt orthogonalisation procedure~\citep{lallemand2000theory}. Explicitly, the transformation matrix can be written as:
\begin{equation}
\begin{bmatrix}
1  &1  &1  &1  &1  &1  &1  &1  &1 \\ 
-4 & -1 &  -1& -1 & -1 &2  & 2 &2  &2 \\ 
 4& -2 &-2  &-2 & -2 &1  &1  &1  & 1\\ 
 0& 1 & 0 &-1 &0  &1 &-1  &-1  &1 \\ 
 0& -2 &0  &2  &0  &1  &-1  &-1  &1 \\ 
 0& 0 &1  &0  &-1  &1  &1  & -1 & -1\\ 
 0& 0 & -2 & 0 & 2 & 1 & 1 &  -1& -1\\ 
 0& 1 & -1 & 1 &  -1& 0 & 0 & 0 &0 \\ 
 0&  0&  0&  0&  0& 1 & 1 & 1 &1 ,\,
\end{bmatrix}
\end{equation}
The collision diagonal matrix $S$ is a diagonal of relaxation rates. The nine eigen values of $S$ are all between 0 and 2 so as to maintain linear stability. The LGBK model is a special case in which the nine relaxation times are all equal and the collision matrix $S=\frac{1}{\tau}I$, where $I$ is the identity matrix. The values of the elements in the collision matrix are: $s_8 = s_9 = \frac{1}{\tau}$ and $s_1 = s_4 = s_6 = 1.0$ and the others vary between 1.0 and 2.0 for linear stability. 

\begin{figure}[htbp]
  \begin{subfigure}{\textwidth}
    \centering
    \includegraphics[width=0.88\linewidth]{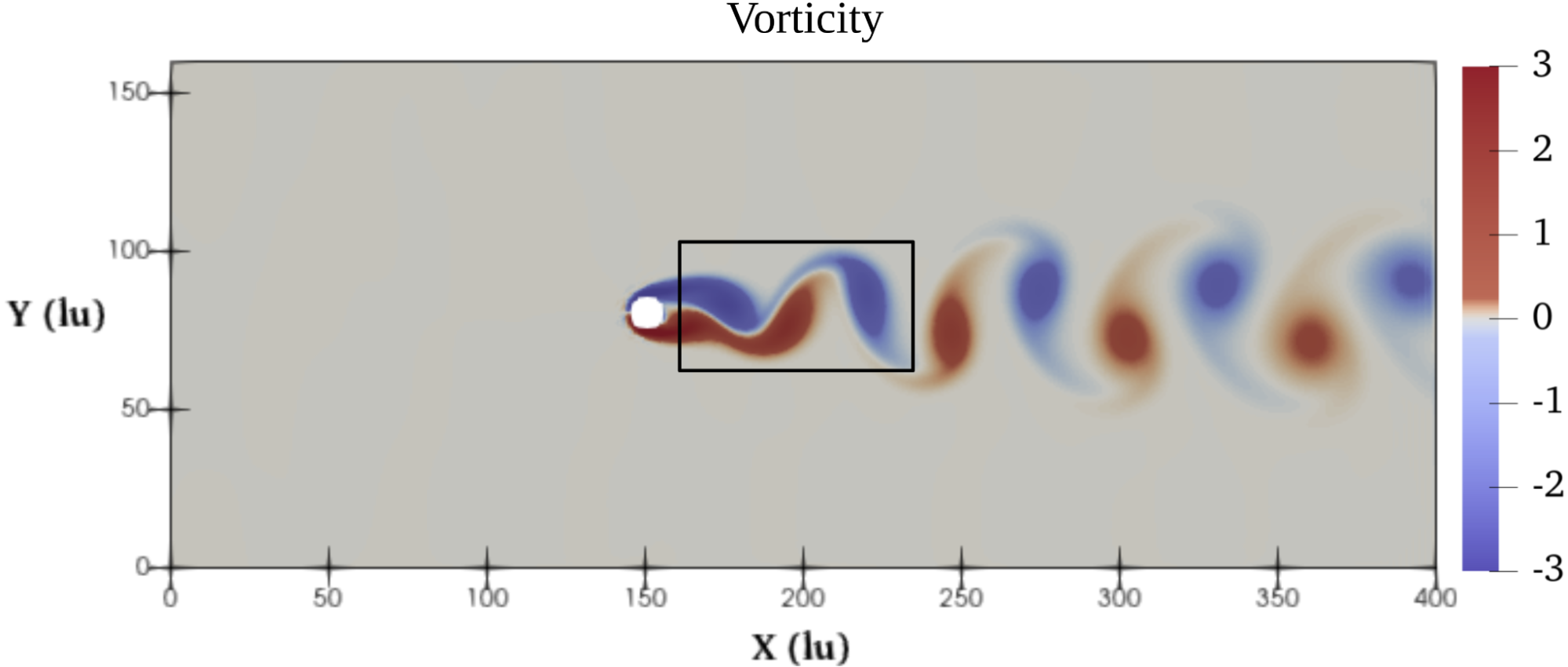}
    \caption{}
  \end{subfigure}%

  \begin{subfigure}{\textwidth}
   \ContinuedFloat
    \centering
    \includegraphics[width=\linewidth]{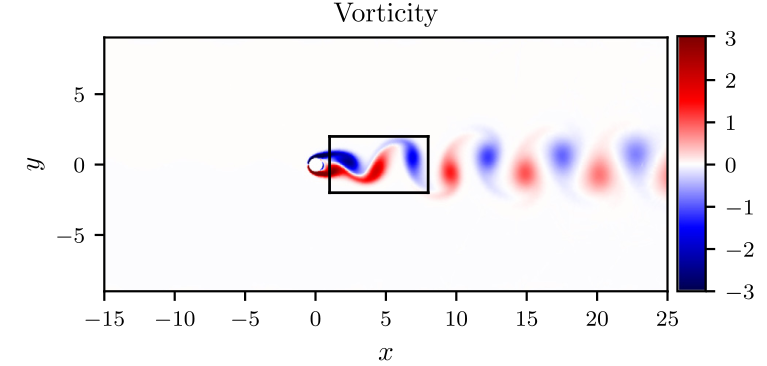}
    \caption{}
  \end{subfigure}%
\caption{Incompressible flow and dynamic vortex shedding past a circular cylinder at $R_e$ = 100 using: a) LBM simulation using MRT collision operator; (b) Navier-Stokes equations~\citep{raissi2019physics}. The region of interest is the rectangle located downstream from the cylinder.}
\label{Fig:shedding}
\end{figure}

\citet{raissi2019physics} investigated a realistic incompressible fluid flow scenario governed by the Navier-Stokes equations. They utilized a physics-informed neural network to identify parameters equivalent to the kinematic viscosity present in the Navier-Stokes equations. Using automatic differentiation, we will evaluate the fluid viscosity similar to Raissi's work. We set a non-dimensional free stream velocity of \( u_\infty = 0.2 \, \text{lu} \cdot \text{(ts)}^{-1} \), a zero-pressure outflow condition on the right boundary situated \( 25 \)-cylinder diameters ( \( D = 10 \, \text{lu} \) ) downstream from the cylinder, and periodicity for both the top and bottom boundaries. We estimate the true kinematic viscosity \( \nu_{\text{LBM}} \) as \( 0.02 \, \text{(lu)}^2 \cdot \text{(ts)}^{-1} \), based on the lattice spacing ( \( \Delta x \) ) and the time step ( \( \Delta t \) ), considering \( u_\infty = 1 \), \( D = 1 \), and \( \nu_{\text{real}} = 0.01 \).

\cref{Fig:shedding} shows the incompressible flow and dynamic vortex shedding around a circular cylinder at $R_e = 100$. The fluid domain has a width of 400 lu and a height of 160 lu, consistent with the dimensions of 40 and 16 in~\citet{raissi2019physics}. We observe the same periodic steady-state behavior marked by an asymmetrical vortex shedding pattern in the cylinder wake, known as the Kármán vortex street~\citep{von2004aerodynamics}. The region of interest is the rectangle located downstream of the cylinder. Our goal is to inversely deduce the kinematic viscosity (with a target value of 0.02 lu$^2$/ts) based on the steady-state velocity distribution in the region of interest.

\cref{Fig:shedding} shows the incompressible flow and dynamic vortex shedding around a circular cylinder at Re = 100. The fluid domain has a width of 400 lu and a height of 160 lu, in consistency with the dimensions of 40 and 16 in~\citet{raissi2019physics}. We observe the same periodic dynamic behavior marked by an asymmetrical vortex shedding pattern around the cylinder, known as the Kármán vortex street~\citep{von2004aerodynamics}. The region of interest is the rectangle located downstream of the cylinder. Our goal is to inversely deduce the kinematic viscosity (with a target value of 0.02 (lu)$^2 \cdot$(ts)$^{-1}$) based on the observed periodic velocity distribution in the region of interest (the rectangle located downstream from the cylinder).

\begin{figure}[htbp]
    \centering
    \includegraphics[width=0.85\columnwidth]{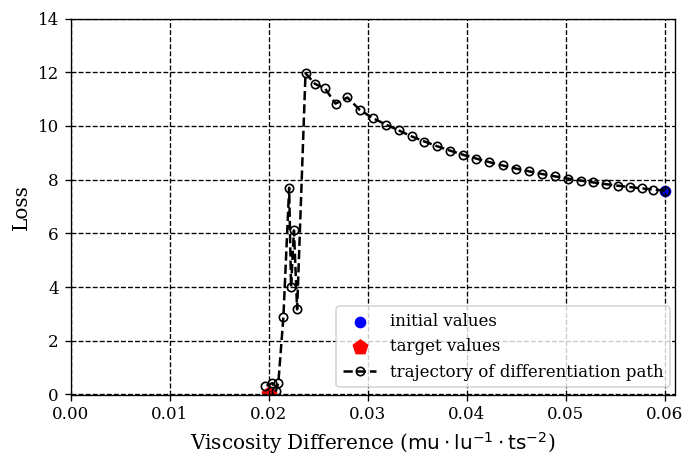}
    \caption{Path of trajectory differentiation for incompressible flow past a circular cylinder at $R_e = 100$ using AD-LBM simulation LBM simulation using MRT collision operator.}
    \label{Fig:viscosity}
\end{figure}

We use a learning rate of \(0.01\) to update the ADAM algorithm for identifying the viscosity by minimizing the flow velocity difference between the target and simulated fields in the target window using AD.~\cref{Fig:viscosity} shows the trajectory of the loss function with viscosity for the AD-LBM simulation of the Kármán vortex street simulation. We successfully identified the viscosity as \(0.022 \, \text{(lu)}^2 \cdot \text{(ts)}^{-1}\) after \(50\) iterations. The loss function initially increased as the viscosity decreased from the initial guess of \(0.06 \, \text{(lu)}^2 \cdot \text{(ts)}^{-1}\) per iteration. However, once the iterative viscosity dropped below \(0.024 \, \text{(lu)}^2 \cdot \text{(ts)}^{-1}\), the loss function sharply decreased from \(12\), indicating the optimized viscosity was being approached. The differentiation path exhibited slight oscillation when converging to the optimized viscosity value. This oscillatory behavior is common in gradient-descent methods and can be attributed to overshooting the minimum due to the discretization of the iterative steps. Although a \(10\%\) error remains, the optimized viscosity of \(0.022 \, \text{(lu)}^2 \cdot \text{(ts)}^{-1}\) can be used as an improved initial guess for subsequent inverse analysis if higher precision is required. The oscillation amplitude could potentially be reduced through smaller step sizes or momentum techniques to smooth the gradient descent trajectory.

\subsection{ Limitations and future research}
Although AD-LBM successfully solves inverse fluid flow problems through granular media, the convergence depends on the initial guess and the gradient at each iteration. As inverse problems are ill-conditioned, developing a good initial guess is critical to achieve successful convergence. A neural network-based initial guess could be coupled directly with AD-LBM to improve optimization and convergence. Further studies could explore using AD-LBM coupled with a differentiable Discrete Element Method in designing and optimizing granular media considering grain movements.

\section{Summary}
This study presents a highly efficient and accurate approach for solving inverse and optimization problems using automatic differentiation. We successfully infer the fluid and granular media's boundary conditions and rheological properties (fluid viscosity or granular permeability) by solving the inverse problems. Implementing the lattice Boltzmann method using automatic differentiation (AD-LBM), we compute the gradients of the results (such as velocities) with respect to the input properties, enabling us to solve inverse and optimization problems involving fluid flow through a granular media. We demonstrate our proposed approach by accurately estimating the pressure profiles in 2D and 3D granular media based on limited target velocities. Additionally, our method identifies fluid viscosity by matching the periodic vorticity in the region of interest. We improve the performance of AD-LBM by implementing GPU-capable differentiable simulators. The AD-LBM method demonstrates considerable advantages in solving inverse and optimization problems involving complex fluid flow in granular media. Further work could incorporate the effects of grain movement into the simulation model, enabling optimization of flow involving dynamic fluid-grain interactions.

\bibliography{literature}  

\onecolumn

\end{document}